\documentclass[amsmath,amssymb,aps,prd]{revtex4-2}

\usepackage{amsmath}
\usepackage{cancel}
\usepackage{ulem}
\usepackage{color}
\usepackage{graphicx}
\usepackage{epsfig}
\usepackage{subfig}
\usepackage{bm}
\usepackage{hyperref}
\usepackage{natbib}
\usepackage{pgfplots,mathtools}
\usepackage{hyperref}
\usepackage{amsmath}
\usepackage{braket}
\usepackage{slashed}
\usepackage[compat=1.0.0]{tikz-feynman}
\usepackage{physics}
\usepackage{xfrac}
\usepackage{algorithm}
\usepackage{lineno}
\usepackage{academicons}
\newcommand{\be}{\begin{equation}}
\newcommand{\ee}{\end{equation}}
\newcommand{\bea}{\begin{eqnarray}}
\newcommand{\eea}{\end{eqnarray}}
\newcommand{\ba}[1]{\begin{array}{#1}}
	\newcommand{\ea}{\end{array}}
\newcommand{\nn}{\nonumber}

\newcommand{\Om}{\Omega}

\newcommand{\del}{\partial}

\newcommand{\la}{\lambda}
\newcommand{\al}{\alpha}
\newcommand{\bt}{\beta}
\newcommand{\Ga}{\Gamma}
\newcommand{\ga}{\gamma}
\newcommand{\D}{\Delta}
\newcommand{\orcid}[1]{\href{https://orcid.org/#1}{\includegraphics[width=8pt]
{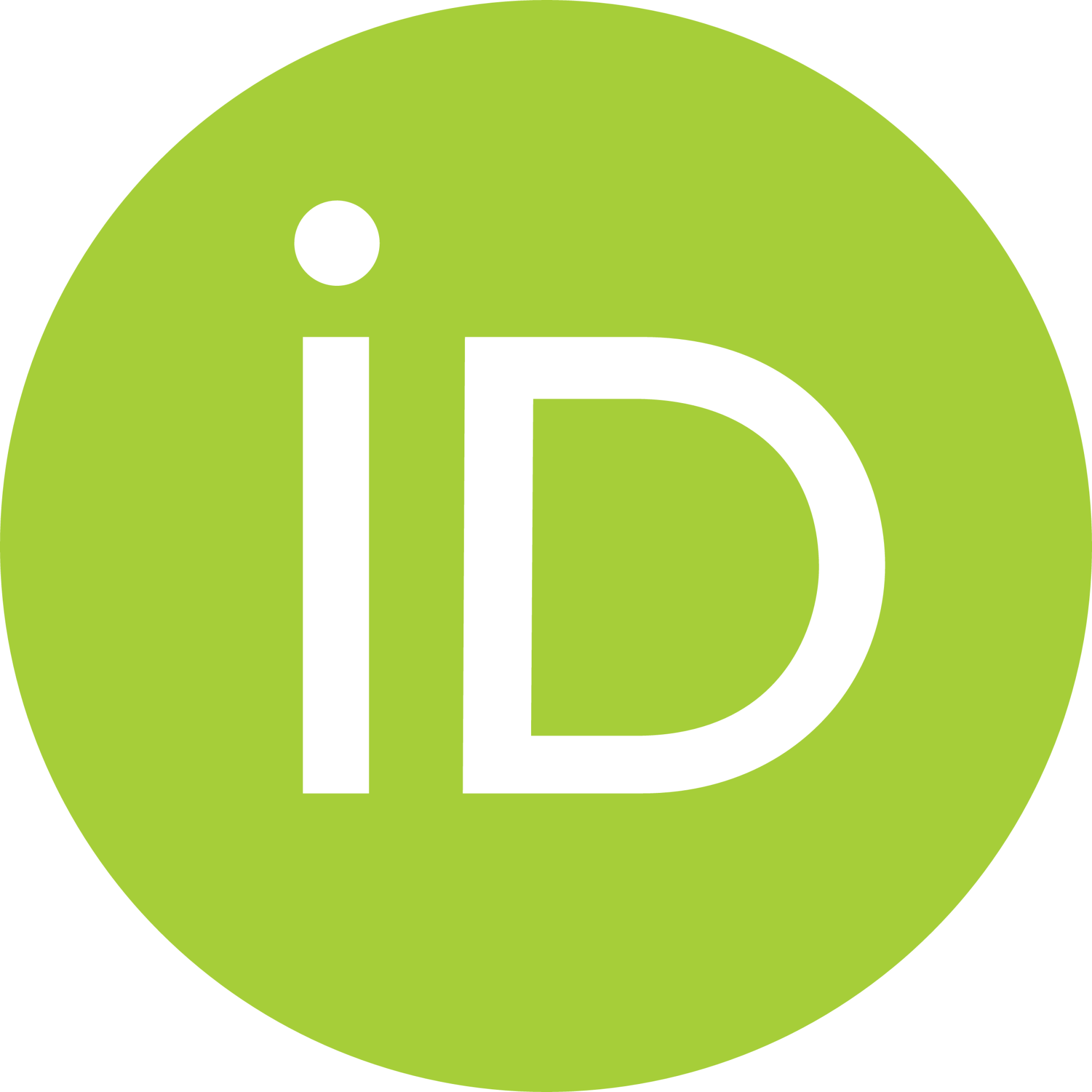}}}
\begin{document}
\title{Effect of Coriolis Force on Electrical Conductivity Tensor for Rotating Hadron Resonance Gas}
\author{Nandita Padhan$^1$, Ashutosh Dwibedi$^2$, Arghya Chatterjee$^1$
 , Sabyasachi Ghosh$^2$}
\affiliation{$^{1}$ Department of Physics, National Institute of Technology Durgapur, Durgapur, 713209, West Bengal, India}
\affiliation{$^2$Department of Physics, Indian Institute of Technology Bhilai, Kutelabhata, Durg, 491001, Chhattisgarh, India}

    \begin{abstract}
     We have investigated the influence of the Coriolis force on the electrical conductivity of hadronic matter formed in relativistic nuclear collisions, employing the Hadron Resonance Gas (HRG) model. 
     A rotating matter in the peripheral heavy ion collisions can be expected from the initial stage of quark matter to late-stage hadronic matter. Present work is focused on rotating hadronic matter, whose medium constituents - hadron resonances can face a non-zero Coriolis force, which can influence the hadronic flow or conductivity. We estimate this conductivity tensor by using the relativistic Boltzmann transport equation.
     In the absence of Coriolis force, an isotropic conductivity tensor for hadronic matter is expected. However, our study finds that the presence of Coriolis force can generate an anisotropic conductivity tensor with three main conductivity components - parallel, perpendicular, and Hall, similar to the effect of Lorentz force at a finite magnetic field. Our study has indicated that a noticeable anisotropy of conductivity tensor can be found within the phenomenological range of angular velocity 
     $\Omega= 0.001-0.02$ GeV and hadronic scattering radius $a=0.2-2$ fm.
    \end{abstract}
\maketitle
\section{Introduction}
In off-central heavy ion collision (HIC) experiments, a large orbital angular momentum (OAM) can be produced, and this initial OAM depends upon factors such as system size, collision geometry, and collision energy, ranging from $10^3\hbar$ to $10^7\hbar$~\cite{STAR:2017ckg,Liang:2004ph,Becattini:2007sr}. After the collision, the spectators carry some of the angular momenta, and the rest is transferred to the produced quark-gluon matter. The initial OAM transferred to the medium is stored in the initial fluid velocity profile of the quark matter and at a later stage in the hadronic matter in the form of local vorticity. This OAM can induce various effects within the medium, like spin polarization, chiral vortical effect (CVE), etc. The vorticity leads to the alignment of hadrons along its direction, influenced by spin-orbit coupling. When considering all space-time points on the freeze-out hypersurface, local polarization accumulates, resulting in a global polarization aligned with the reaction plane or the angular momentum of the colliding nuclei. The global polarization of $\Lambda$ and $\Bar{\Lambda}$ particles has been measured by the STAR Collaboration in Au+Au collisions across a range of collision energies ($\sqrt{s_{NN}}=$7.7-200 GeV), revealing a decreasing trend with collision energies~\cite{STAR:2017ckg}. Moreover, in the recent study with improved statistics at $\sqrt{s_{NN}}=$ 200 GeV, a polarization dependence on the event-by-event charge asymmetry was observed, indicating a potential contribution to the global polarization from the axial current induced by the initial magnetic field~\cite{STAR:2018gyt}. Additionally, spin alignment has also been observed in vector mesons, with recent measurements conducted at RHIC and LHC further contributing to our understanding of spin phenomena in heavy ion collisions~\cite{ALICE:2019aid,STAR:2022fan,ALICE:2022dyy}. 

Now, from the theoretical direction, the effect of large OAM on the medium constituents has been studied even before the experimental work of STAR Collaboration~\cite{STAR:2017ckg} in the Refs.~\cite{Liang:2004ph,Becattini:2007sr,BECATTINI20082452,Becattini:2013fla,Becattini:2013vja,Betz:2007kg,LIANG200520,Wang_2008,XuGuangHuang2011}. The experimental finding of global polarization of $\Lambda$ and $\Bar{\Lambda}$ particles in $2017$ ~\cite{STAR:2017ckg} also stimulated many theoretical investigations of vorticity and spin polarization effects in HIC~\cite{Becattini:2021suc,Xia:2018tes,Florkowski2018,Florkowski_2018,Florkowski:2019qdp,XuGuangHuang2019,Wu:2019yiz,XuGuangHuang:2020dtn,Fu:2021pok,XuGuangHuang_2022,XuGuangHuang2022}. The study of vorticity and the polarization of the particles produced by HICs has been done by multitude of theoretical approaches. The references~\cite{Wang_2012, Wang_2013,Wang_2016, Wang_2017,GAO2015542,Yang2017,Gao2017,Gao2018,Liao2018,Gao2019,Gao_2019,Hattori2019,Zhuang_2019M,Rischke2019,yang2020effective,Rischke2021w,Rischke_2021Nk}, with the help of covariant Wigner functions and quantum kinetic equations, have described the chiral effects and the spin polarization of final state particles. On the other hand, the authors of Refs.~\cite{becattini2007microcanonical, Becattini:2007sr, BECATTINI20082452, BECATTINI20101566, Becattini:2013fla, Becattini:2013vja, Becattini:2015ska, Becattini:2021suc} used the theory of relativistic statistical mechanics for a plasma in global equilibrium under rotation to describe the polarization of particles emitted from the kinetic freeze-out hypersurface. In contrast, in Refs.~\cite{Betz:2007kg, Liang:2004ph, LIANG200520, Wang_2008,chen2009general, XuGuangHuang2011}, the spin-orbit interaction in QCD has been used to describe the transfer of initial OAM density into the spin angular momentum, ultimately resulting in the spin polarization of particles. Moreover, the authors of Refs.~\cite{Becattini2011,Florkowski2018,Florkowski_2018,Florkowski:2018fap,Florkowski:2018ahw,BECATTINI2019419,Florkowski:2019qdp,BHADURY2021,Bhadury_2021,daher2022equivalence,Bhadury2022} have developed a kinetic framework to establish the equations of spin hydrodynamics by including the spin tensor. In addition, several transport and hydrodynamical models ~\cite{XuGuangHuang2016,Jiang2016,Pang:2016igs,Li:2017slc,Xia:2018tes,XuGuangHuang2019,XuGuangHuang:2019xyr,Wu:2019yiz,XuGuangHuang:2020dtn,Fu:2021pok,XuGuangHuang_2022,XuGuangHuang2022} have also been used to estimate spin polarization and vorticity results in HIC quantitatively. Thermodynamics of the hadronic medium under rotation have recently been explored by Refs.~\cite{Pradhan:2023rvf, Sahoo:2023xnu, Mukherjee:2023ijv}. The phase structure of rigidly rotating plasma has been explored in Refs.~\cite{Chernodub:2020qah,Chernodub:2017ref}. The Lattice Quantum Chromodynamics (LQCD) calculations in the presence of rotations can be found in the Refs.~\cite{Braguta:2023yjn,Chernodub:2022veq}.

There is a similarity between magnetic field and rotation. The picture of Lorentz force in the presence of magnetic fields is quite similar to the picture of the Coriolis force in the presence of rotation. In Refs.~\cite{J_Sivardiere_1983, Johnson2000-px,Sakurai1980}, the equivalence between the Coriolis force and Lorentz force has been explored. In the presence of magnetic fields, the transport coefficients of the systems become an-isotropic\cite{Dey:2019axu,Dash:2020vxk,Dey:2020awu,Ghosh:2019ubc,Dey:2019vkn,Kalikotay:2020snc,Dey:2021fbo,Satapathy:2021cjp,Das:2019wjg,Das:2019ppb,Dey:2019axu,Chatterjee:2019nld,Hattori:2016cnt,Hattori:2016lqx,Satapathy:2021wex}. Accordingly, one should expect the anisotropic structure of the transport coefficients in a rotating frame due to the effect of the Coriolis force.
In this paper, we will show how the electrical conductivity of the rotating hadronic matter formed in HIC can be modified in the presence of Coriolis force. In the papers ~\cite{Aung:2023pjf,Dwibedi:2023akm}, the authors have described how a rotating medium's shear viscosity and electrical conductivity become anisotropic in the presence of Coriolis force. Here, we have extended the formalism of Ref.~\cite{Dwibedi:2023akm} from non-relativistic to the relativistic case, which is applicable to calculate the electrical conductivity of the rotating hadronic matter. To fulfill this purpose, we employed the covariant Boltzmann transport equation (BTE) under the rotational background. In this framework, the space-dependent metric makes the connection coefficients non-zero, which in turn manifest themselves as the pseudo force terms in the covariant Boltzmann equation. To simplify the analysis, we ignored the quadratic and higher power of angular velocity $\Om$ present in the covariant BTE. This simplification leads to the elimination of centrifugal effects and makes a seamless comparison possible with the magnetic field scenario.
We modeled our rotating hadronic medium by resorting to the popular HRG model.  
This model is founded upon principles derived from statistical mechanics of multi-hadron species. Using S-matrix calculation, it has been shown that in the presence of narrow resonances, the thermodynamics of the interacting gas of hadrons can be approximated by the ideal gas of hadrons and its resonances~\cite{Dashen:1969ep,Dashen:1974jw}.  
The HRG model has been extensively used to study thermodynamics~\cite{Karsch:2003zq,Braun-Munzinger:2015hba} and conserved charge fluctuations~\cite{Begun:2006jf,Nahrgang:2014fza,HotQCD:2012fhj,Bhattacharyya:2013oya,Chatterjee:2016mve}, as well as transport coefficients~\cite{Gorenstein:2007mw,Noronha-Hostler:2012ycm,Tiwari:2011km,Pradhan:2022gbm,Noronha-Hostler:2008kkf,Kadam:2014cua,Zhang:2019uct,Samanta:2017ohm,Ghosh:2019fpx,Ghosh:2018nqi,Rocha:2024rce}, which are quite accepted for heavy ion collision phenomenology. 
Recently, Refs.~\cite{Dash:2020vxk,Dey:2020awu,Das:2019wjg,Das:2019ppb} have demonstrated the role of Lorentz force in creating anisotropic transportation of
HRG system. However, the role of the Coriolis force in creating a similar kind of anisotropic transportation for the HRG system has not been studied yet, and
here, we are first time going for this kind of investigation.\\

The article is arranged as follows: in Sec.~\ref{sec:Formalism}, we develop the necessary formalism needed for calculating electrical conductivity tensor in the presence of rotation. The master formula for hadronic matter with the hadron resonance gas model (HRG) and  QGP with massless approximation is provided in \ref{hrg} and Sec.~\ref{mqgp} from which the results are generated.
In Sec.~\ref{Sec:Results_Discussion}, we present the numerical results with the plots of the variation of conductivity for QGP and hadronic matter both in the presence and absence of rotation. The article is summarised in Sec.~\ref{Sec:Summary}.

\section{Formalism}
\label{sec:Formalism}
To begin with, let us briefly recapitulate the non-relativistic kinetic model used to calculate the shear viscosity and electrical conductivity of the rotating nuclear matter in Refs.~\cite{Aung:2023pjf,Dwibedi:2023akm}. The non-relativistic kinetic model consists of implementing the rotating coordinate transformation to break the particle velocity into two parts: (1) velocity $\vec{v}_{r}$ seen from a frame rotating with angular velocity $\vec{\Om}$, and (2) a rigid rotor velocity $\vec{\Om}\times \vec{r}$. Then, for the time evolution of the distribution function, the BTE is written down in the rotating frame (coordinates) with apparent or pseudo forces entering as the force terms. The transport properties are studied by expanding the solution to the BTE around the local equilibrium distribution in the rotating frame. In generalizing this model to the relativistic scenario, we will keep the same physical picture as that of its non-relativistic counterpart. To carry out the relativistic extension in practice, we will obtain the equation of motion (EOM) and covariant BTE in a rotating frame with the help of rotating frame metric tensor $g_{\mu\nu}(\text{not same as the flat metric } \eta_{\mu\nu})$ and connection coefficients $\Gamma_{\mu \lambda}^{\alpha}$ (vanishes in an inertial frame). As our relativistic kinetic model of rotating nuclear matter, let us consider a rotating system of hadrons moving with the momenta $\vec{p}_{r}$ (subscript $r$ stands for hadron species; it should not be confused with radial coordinate).

The micro- and macroscopic expressions of current density for these collections of hadron resonances under an applied electric field $\Vec{\Tilde{E}}\equiv \Tilde{E}\hat{e}$ are,
\begin{eqnarray}
    J^{i} &=&\sum\limits_{r}J_{r}^{i}= \sum\limits_{r} g_{r} q_{r} \int{\frac{d^{3}\vec{p_r}}{{(2\pi})^{3}}} \frac{p^{i}_{r}}{E_r} \delta{f_{r}}, (p_{r0}\equiv E_{r})\label{A1}\\
   J^{i}&=&\sum\limits_{r}J_{r}^{i}=\sum\limits_{r}\sigma_{r}^{ij}\Tilde{E}_{j}\equiv\sigma^{ij}\Tilde{E}_{j},\label{ma1}
\end{eqnarray}
where $r$ is the label characterizing the different hadrons and resonances with charge $q_{r}$, energy $E_r$, momentum $p_r$, and degeneracy $g_{r}$. The $\delta{f_{r}}$ quantifies the deviation of the system from local equilibrium. $J^{i} =\sum\limits_{r}J_{r}^{i}$ and $\sigma^{ij}=\sum\limits_{r}\sigma_{r}^{ij}$ are respectively the current density and conductivity tensor due to all the hadronic species comprised of baryons and mesons. The microscopic expression of current density in the HRG phase provided in Eq.~(\ref{A1}) can be compared with the macroscopic expression $J^{i}=\sigma^{ij}\Tilde{E}_{j}$ to obtain the conductivity tensor $\sigma^{ij}$. The deviation function $\delta f_{r}$ written in Eq.~(\ref{A1}) corresponds to the difference between the total distribution function and equilibrium distribution function for the $r$th hadronic species, i.e., $\delta f_{r}=f_{r}-f_{r}^{0}$.
We will assume that the system is slightly out of equilibrium so that $\delta f_{r}$ can be treated as a perturbation. The perturbation $\delta f_{r}$ can be determined by using the Boltzmann transport equation (BTE) in a rotating frame. 
Before writing the BTE in the rotating frame, we will briefly address the rotating frame transformation and equation of motion of a hadron in a rotating frame. To analyze the hadronic medium rotating with an angular velocity $\Vec{\Omega}\equiv \Omega \hat{\omega}$, we will make an explicit coordinate transformation from the inertial coordinates: $(t,x,y,z)$ to the co-rotating coordinates: $(t^{\prime},x^{\prime},y^{\prime},z^{\prime})$ as follows\cite{Chernodub:2016kxh,Chernodub:2017ref,Chernodub:2020qah,FUJIMOTO2021136184}:
\be
\begin{rcases}
t^{\prime}&=t\\
x^{\prime}&=x~cos~\Om t + y~sin~\Om t\\
y^{\prime}&=-x~sin~\Om t + y~cos~\Om t\\
z^{\prime}&=z\\
\end{rcases}\label{R1}~,
\ee
where we assumed the angular velocity to be in the z-direction, i.e., $ \hat{\omega}= \hat{k}$. The squared differential length element $ds^{2}$ and the co-rotating frame metric $g^{\mu\nu}$ can be obtained from the Eq.~\ref{R1} as\cite{Ebihara:2016fwa,Chernodub:2017mvp,Kapusta:2019sad}:
\bea
&&ds^{2}= g_{\mu\nu} dx^{\prime\mu}dx^{\prime\nu}=dt^{\prime^{2}}(1-\Om^2x^{\prime^2}-\Om^2y^{\prime^2})+2\Om y^{\prime}dt^{\prime}dx^{\prime}-2\Om x^{\prime} dt^{\prime}dy^{\prime}-dx^{\prime^2}-dy^{\prime^2}-dz^{\prime^2},\nn\\
&& g_{\mu\nu}=
\begin{pmatrix}
      1-\Om^2x^{\prime^2}-\Om^2y^{\prime^2}  & \Om y^{\prime} & -\Om x^{\prime} & 0\\
    \Om y^{\prime}                           &      -1        &         0       & 0\\
   -\Om x^{\prime}                           &       0        &        -1       & 0 \\
    0                                        &       0        &         0       & -1 
    \label{R2}
\end{pmatrix}.
\eea
The connection coefficients $\Gamma^{\al}_{\mu\la}$ in  co-rotating coordinates can be expressed in terms of the derivative of the metric tensor\cite{misner2017gravitation,schutz2009first,Cercignani200211},
\bea
&&\Gamma_{\mu \lambda}^{\alpha}=\frac{1}{2}g^{\alpha \nu}\left(\frac{\del g_{\nu \mu}}{\del x^{\lambda}} +\frac{\del g_{\lambda \nu}}{\del x^{\mu}} - \frac{\del g_{\mu \lambda}}{\del x^{\nu}}\right),\label{R3}
\eea
where we dropped the overhead primes from the rotating frame variables. Since all the subsequent calculations will be performed from the rotating frame, we will drop the rotating frame quantities' overhead primes to simplify our notation.
Upon explicit calculation with the help of Eq.~\ref{R3}, one finds that out of $64$ components of $\Gamma^{\al}_{\mu\la}$ only six are non-zero\cite{Kapusta:2019sad}: $\Gamma_{00}^{1}=-\Om^{2}x, \Gamma_{00}^{2}=-\Om^{2}y, \Gamma_{20}^{1}=\Gamma_{02}^{1}=-\Om, \Gamma_{10}^{2}=\Gamma_{01}^{2}=\Om$.
One can write down the equation of motion (EOM) of a hadron in the rotating frame with the help of connection coefficients $\Gamma^{\al}_{\mu\la}$ as\cite{Cercignani200210,misner2017gravitation,schutz2009first},
\be
\frac{d p^{\al}}{d\tau} + \frac{1}{m}  p^{\mu}p^{\la} \Ga_{\mu \la}^{\al} =F^{\al},\label{R4}
\ee
where  $p^{\al}$, $F^{\al}$, and $\tau$ are the four-momentum, four-force, and proper time, respectively. To see the similarity between the EOM provided by  Eq.~\ref{R4} with the classical non-relativistic EOM in the rotating frame~\cite{goldstein2011classical,kleppner2014introduction}, we rewrite Eq.~\ref{R4} with the substitution of connection coefficients,
\bea
&&\frac{d\Vec{p}}{dt}=\ga_{v} m(\Vec{\Om}\times\Vec{r})\times \Vec{\Om}+ 2\ga_{v} m(\Vec{v}\times\Vec{\Om}),\label{R5}
\eea
where we employed $\ga_{v}=\frac{dt}{d\tau}$, $p^{\al}=(\ga_{v} m,\ga_{v} m \Vec{v})=(\ga_{v} m,\Vec{p})$ and $F^{\al}=0$. In comparison with the classical non-relativistic EOM \cite{goldstein2011classical,kleppner2014introduction}, we noticed that the first and second terms of the RHS of Eq.~\ref{R5} correspond to centrifugal and Coriolis force with the extra multiplicative Lorentz factor $\ga_{v}$. Moreover, comparing the Coriolis force expression $\Vec{F}_{\rm Cor}=2\ga_{v} m(\Vec{v}\times\Vec{\Om})$ established in Eq.~\ref{R5} with the expression of Lorentz force $\Vec{F}_{\rm Lor}=q (\Vec{v}\times\Vec{B})$, a direct equivalence can be obtained between the $\Vec{\Om}$ in the rotating frame and magnetic field $\Vec{B}$ in the inertial frame. In the rotating frame with the metric tensor $g_{\mu\nu}$ the Lorentz factor $\ga_{v}$ can be expressed as \cite{Cercignani200211,Kremer:2012nk}:
\bea
&& \ga_{v}=\frac{1}{\sqrt{g_{00}(1+\frac{g_{0i}v^{i}}{g_
{00}})^{2}-v^{2}}},\label{RR1}
\eea
where we used the definitions $v^{i}\equiv \frac{dx^{i}}{dt}$,   $v^{2}\equiv(\frac{g_{0i}g_{0j}}{g_{00}}-g_{ij})v^{i}v^{j}$. Similarly, one can easily show by using the relation $p^{\mu}p_{\mu}=p^{\mu}p^{\nu}g_{\mu\nu}=m^{2}$ and $p^{0}=\frac{p_{0}-g_{0i}p^{i}}{g_{00}}$ that\cite{Cercignani200212,Kremer:2012dd,Kremer:2012nk,Kremer:2013fxa,doi:10.1142/S0219887814600056,PhysRevE.91.052139},
\bea
p_{0}=E=\sqrt{m^{2}g_{00}+(g_{0i}g_{0j}-g_{00}g_{ij})p^{i}p^{j}}.\label{RR2}
\eea
Now, we will write down the BTE in the co-rotating frame variables by equating the variation of distribution function $f_{r}(x^{\mu},\Vec{p}_{r})$ along the hadron's world line with the collision kernel $C[f_{r}]$\cite{Cercignani200212,DEBBASCH20091079,DEBBASCH20091818}:
\bea
&&m_{r}\frac{df_{r}(x^{\mu},\Vec{p}_{r})}{d\tau}=C[f_{r}]\nn\\
\implies &&m_{r} \frac{dx^{\mu}}{d\tau}\frac{\del f_{r}}{\del x^{\mu}}+ m_{r} \frac{dp^{i}_{r}}{d\tau}\frac{\partial f_{r}}{\partial p^{i}_{r}} =C[f_{r}]\nn\\
\implies &&p^{\mu}_{r}\frac{\del f_{r}}{\partial x^{\mu}}- \Ga_{\mu \lambda}^{i} p^{\mu}_{r}p^{\la}_{r} \frac{\partial f_{r}}{\partial p^{i}_{r}}+m_{r} F^{i}_{r}\frac{\partial f_{r}}{\partial p^{i}_{r}}= C[f_{r}]~,\label{R6}
\eea
where we used the EOM provided in Eq.~\ref{R4} to get the last step. The Eq.~\ref{R6} can be easily written in a fully covariant manner as\cite{Romatschke:2011qp,Cercignani200212,DEBBASCH20091079,DEBBASCH20091818}:
\bea
&&p^{\mu}_{r}\frac{\del f_{r}}{\partial x^{\mu}}- \Ga_{\mu \lambda}^{\al} p^{\mu}_{r}p^{\la}_{r} \frac{\partial f_{r}}{\partial p^{\al}_{r}}+m_{r} F^{\al}_{r}\frac{\partial f_{r}}{\partial p^{\al}_{r}}= C[f_{r}],\label{R7}
\eea
where we treat $f_{r}$ to be function of independent variables $x^{\al}$ and $p^{\al}_{r}$, i.e., $f_{r}=f_{r}(x^{\al},p^{\al})$. The four force $F^{\al}$ is the electromagnetic force, i.e., $F^{\al}_{r}=q_{r}F^{\al\bt}p_{r\bt}/m_{r}$, where $F^{\al\bt}$ is the faraday tensor. In the above, we provided the covariant form of BTE, paying special attention to rotating coordinate transformation. Nevertheless, the arguments leading to the final form of BTE given in Eq.~\ref{R6} and \ref{R7} can be easily generalized for general coordinate transformation and gravity, with only notable changes appearing in the explicit expression metric tensor and connection coefficients\cite{Romatschke:2011qp,Cercignani200212,DEBBASCH20091079,DEBBASCH20091818}. The solution of Eq.~\ref{R7} can be obtained by assuming the system to be slightly out of equilibrium so that we can write $f_{r}=f^{0}_{r}+\delta f_{r}$ with $f^{0}_{r}=1/[e^{(p^{\al}_{r}u^{\bt}g_
{\al\bt}-\mu)/T}-a]=1/[e^{(p^{\al}_{r}u_{\al}-\mu)/T}-a]$, where $a=-1$ for baryons and $a=+1$ for mesons. The fluid velocity $u^{\mu}$ occurring in the equilibrium distributions $f^{0}_{r}$ can be written as $u^{\mu}=\ga_{u}
(1,\Vec{u})$,where 
\bea
&& \ga_{u}=\frac{1}{\sqrt{g_{00}(1+\frac{g_{0i}u^{i}}{g_
{00}})^{2}-u^{2}}}, (u^{2}\equiv(\frac{g_{0i}g_{0j}}{g_{00}}-g_{ij})u^{i}u^{j}).\label{RR3}
\eea
To simplify the analysis, we will approximate the collision term $C[f_{r}]$ by the expression given by relaxation time approximation (RTA), i.e., $C[f_{r}]=-(u^{\al}p_{\al})\frac{f_{r}-f^{0}_{r}}{\tau_{c}}$. Substituting $f_{r}=f^{0}_{r}+\delta f_{r}$ in Eq.~\ref{R7} we have,
\bea
&&p^{\mu}_{r}\frac{\del f^{0}_{r}}{\partial x^{\mu}}- \Ga_{\mu \lambda}^{\al} p^{\mu}_{r}p^{\la}_{r} \frac{\partial}{\partial p^{\al}_{r}} (f^{0}_{r} +\delta f_{r})+q_{r}F^{\al\bt}p_{r\bt}\frac{\partial }{\partial p^{\al}_{r}}(f^{0}_{r}+\delta f_{r})= C[f_{r}]=-(u^{\al}p_{\al})\frac{\delta f_{r}}{\tau_{c}},\label{R8}
\eea
where we neglected space-time gradient of $\delta f_{r}$ in the first term of Eq.~\ref{R8} since it give rise to second order gradient effects. The faraday tensor $F^{\mu\nu}$ can be decomposed into electric $\tilde{E}^{\mu}$ and magnetic part $B^{\mu\nu}$ with the help of fluid velocity $u^{\mu}$ and projector $\D^{\mu\nu}=g^{\mu\nu}-u^{\mu}u^{\nu}$. We can write $F^{\mu\nu}=\tilde{E}^{\mu}u^{\nu}-\tilde{E}^{\nu}u^{\mu}+B^{\mu\nu}$, where $\tilde{E}^{\mu}\equiv F^{\mu\nu}u_{\nu}$ and $B^{\mu\nu}\equiv \D^{\mu}_{\al}F^{\al\bt}\D^{\nu}_{\bt}$. In the present article, we will focus only on the effect of the electric field; therefore, we will ignore the magnetic part $B^{\mu\nu}$ and rewrite Eq.~\ref{R8} as
\bea
&&\bigg[p^{\mu}_{r}\frac{\del f^{0}_{r}}{\partial x^{\mu}}- \Ga_{\mu \lambda}^{\al} p^{\mu}_{r}p^{\la}_{r} \frac{\partial f^{0}_{r}}{\partial p^{\al}_{r}}  +q_{r}p_{r\bt}(\tilde{E}^{\al}u^{\bt}-\tilde{E}^{\bt}u^{\al})\frac{\partial f^{0}_{r} }{\partial  p^{\al}_{r}}\bigg]- \Ga_{\mu \lambda}^{\al} p^{\mu}_{r}p^{\la}_{r} \frac{\del\delta f_{r}}{\partial p^{\al}_{r}} = -(u^{\al}p_{\al})\frac{\delta f_{r}}{\tau_{c}}\nn\\
\implies && -f^{0}_{r}(1+af^{0}_{r})\bigg[\frac{p^{\mu}_{r}p^{\al}_{r}}{T}(\del_{\mu}u_{\al}-\Ga^{\sigma}_{\mu\al}u_{\sigma})+p^{\mu}_{r}(u_{\al}p^{\al}_{r})\del_{\mu}\frac{1}{T}-p^{\mu}_{r}\del_{\mu}\frac{\mu}{T}-\frac{q_{r}\tilde{E}_{\nu}p^{\nu}_{r}}{T}\bigg]\nn\\
&&-\Ga^{\sigma}_{\mu\la}p^{\mu}p^{\la}\frac{\del \delta f_{r}}{\del p^{\sigma}_{r}}= -(u^{\al}p_{\al})\frac{\delta f_{r}}{\tau_{c}}. \label{R9}
\eea
The first three terms in the square bracket of Eq.~\ref{R9} give rise to viscous stresses and diffusion; they are related to the shear viscosity, bulk viscosity, and thermal conductivity of the hadronic medium. Since the present article is planned to calculate electrical conductivity, we will focus on the dissipative flow arising from the electric field. For the calculation of electrical conductivity Eq.~\ref{R9} can be rewritten as,
\bea
\frac{q_{r}f^{0}_{r}(1+af^{0}_{r})}{T} \tilde{E}_{\mu}p^{\mu}-\Ga^{\sigma}_{\mu\la}p^{\mu}p^{\la}\frac{\del \delta f_{r}}{\del p^{\sigma}_{r}}=-(u^{\al}p_{\al})\frac{\delta f_{r}}{\tau_{c}}.\label{R10}
\eea
The explicit form of $p_{0}$ in the rotating frame can be written with the use of Eq.~\ref{R2} in Eq.~\ref{RR2} as:
\bea
&&p_{0}=E=\sqrt{m^{2}(1-\Om^{2}\rho^{2})+(1-\Om^{2}x^{2})(p^{1})^{2}+ (1-\Om^{2}y^{2})(p^{2})^{2} + (1-\Om^{2}\rho^{2})(p^{3})^{2}-2\Om^{2}xy~p^{1}p^{2}},~(\rho\equiv x^{2}+y^{2}).\label{RF1}
\eea
Similarly, the contravariant time component of the momentum vector $p^{0}$ is given by,
\bea
&&p^{0}=\frac{p_{0}+\vec{r}\cdot(\vec{p}\times\vec{\Om})}{g_{00}}.\label{RF2}
\eea
The invariant momentum space measure $dP$ is given by,
\bea
&& dP\equiv \sqrt{g}\frac{d^{3}p}{p_{0}}, \text{ where } g\equiv-det((g_{\mu\nu})).\label{RF3} 
\eea
In the rotating frame, $det((g_{\mu\nu}))=-1$ and $dP=\frac{d^{3}p}{p_{0}}=\frac{d^{3}p}{E}$, which justifies the invariant momentum measure in our original definition in Eq.~\ref{A1}. At this juncture, it should be pointed out that, in principle, one can solve Eq.~\ref{R10} to obtain the electrical conductivity of rotating HRG without resorting to any further approximation. Nevertheless, the result becomes algebraically convoluted because of the space-dependent metric $g_{\mu\nu}$, which significantly differs from the flat metric $\eta_{\mu\nu}$. As the beginning level of calculation for our relativistic kinetic model of rotating matter, we will rely on two different approximations. These two approximations will simplify our result and make a seamless comparison with the magnetic field scenario possible. The first phase of approximation assumes that globally rotating nuclear matter is produced in HIC by ignoring radial expansion. This implies the fluid velocity $u^{\mu}=\gamma_{u}(1,\Vec{u})\xrightarrow{u^{i}=0}(\frac{1}{\sqrt{g_{00}}},0)$,  in the globally co-moving frame. Similarly, the local equilibrium distribution and electric field can be written as  $f^{0}_{r}=1/[e^{(p_{r\al}u^{\al}-\mu)/T}-a]\xrightarrow{u^{i}=0}1/[e^{(p_{r0}/\sqrt{g_{00}}-\mu)/T}-a]$ and $\tilde{E}_{\nu}=F_{\nu\mu}u^{\mu}\xrightarrow{u^{i}=0}F_{\nu0}/\sqrt{g_{00}}=-\tilde{E}^{i}/\sqrt{g_{00}}$. Eq.~\ref{R10} in the co-moving frame becomes,
\bea
&& -\frac{f^{0}(1+af^{0})}{T}\frac{q}{\sqrt{g_{00}}}\Vec{\tilde{E}}\cdot \Vec{p}+ 2p^{0}(\vec{p}\times\Vec{\Om})\cdot \frac{\del \delta f}{\del \Vec{p}}+(p^{0})^{2}((\Vec{\Om}\times \vec{r})
\times \vec{\Om})\cdot \frac{\del \delta f}{\del \Vec{p}}=-\frac{p_{0}}{\sqrt{g_{00}}}\frac{\delta f}{\tau_{c}}\nn\\
\implies && -\frac{f^{0}(1+af^{0})}{T}\frac{q \Vec{\tilde{E}}\cdot \Vec{p}}{p_{0}}+ \frac{2p^{0}\sqrt{g_{00}}(\vec{p}\times\Vec{\Om})}{p_{0}}\cdot \frac{\del \delta f}{\del \Vec{p}}+\frac{(p^{0})^{2}\sqrt{g_{00}}}{p_{0}}((\Vec{\Om}\times \vec{r})
\times \vec{\Om})\cdot \frac{\del \delta f}{\del \Vec{p}}=-\frac{\delta f}{\tau_{c}},\label{R11}
\eea
where we have suppressed the index $r$ in writing Eq.~\ref{R11}, which will be retained during the calculation of total conductivity. The second phase of approximations will be implemented in solving Eq.~\ref{R11} with the help of the ansatz  $\delta f=-\Vec{p}\cdot\vec{X} \big( \frac{\partial f^0}{\partial E} \big)=\Vec{p}\cdot\vec{X}\frac{f^{0}(1+af^{0})}{T}$, where $\vec{X}$ is still to be determined. This approximation includes the following:  
$\Om x$, $\Om y$, and $\frac{\Om}{T}$ are small, so one can ignore the second or higher order terms containing products of any two of them. Neglecting the quadratic terms in $\Om x$ and/or  $\Om y$ in Eq.~\ref{RF1} we have $p_{0}=\sqrt{m^{2}+\vec{p}^{2}}$. Employing the above approximation, the Coriolis force term in Eq.~\ref{R11} can be written as,
\bea
&& \frac{2p^{0}\sqrt{g_{00}}(\vec{p}\times\Vec{\Om})}{p_{0}}\cdot \frac{\del \delta f}{\del \Vec{p}}=\frac{2}{p_{0}\sqrt{g_{00}}} [p_{0}+\vec{r}\cdot(\vec{p}\times\vec{\Om})](\vec{p}\times \vec{\Om})\cdot\frac{\del \delta f}{\del \Vec{p}}\nn\\
\implies && \frac{2p^{0}\sqrt{g_{00}}(\vec{p}\times\Vec{\Om})}{p_{0}}\cdot \frac{\del \delta f}{\del \Vec{p}}\approx 2 (\vec{p}\times\vec{\Om})\cdot \frac{\del \delta f}{\del \Vec{p}}, (g_{00}=1-\Om(x^{2}+y^{2})\approx 1).\label{RF4}
\eea
Similarly, it can be easily checked that the centrifugal force term $\frac{(p^{0})^{2}\sqrt{g_{00}}}{p_{0}}((\Vec{\Om}\times \vec{r})
\times \vec{\Om})\cdot \frac{\del \delta f}{\del \Vec{p}}$ completely drops out, leaving the following equation to be solved,
\bea
&&\frac{\partial f^0}{\partial E}\frac{\Vec{p}}{p_{0}}\cdot(q\Vec{\Tilde{E}})+2m\gamma_v(\Vec{v}\times\Vec{\Omega})\cdot\frac{\partial\delta f}{\partial\Vec{p}}=-\frac{\delta f}{\tau_{c}}.\label{R12}
\eea

 One has to solve Eq.~\ref{R12} for $\delta f$ to determine HRG current density $J^{i}$ from Eq.~\ref{A1}. So everything boils down to the determination of $\vec{X}$ in the ansatz: $\delta f=-\Vec{p}\cdot\vec{X} \big( \frac{\partial f^0}{\partial E} \big)$. Our system of rotating HRG is no longer isotropic because of the presence of the angular velocity vector $\vec{\Omega}=\Omega\hat{\omega}$. We have two unit vectors $\hat{\omega}$ and $\hat{e}$ in our hand, which can be used to construct another unit vector $\hat{e}\times \hat{\omega}$ perpendicular to both $\hat{\omega}$ and $\hat{e}$. In general, the current density in rotating HRG can have components along $\hat{\omega}$, $\hat{e}$, and $\hat{e}\times \hat{\omega}$. Since the vector $\Vec{X}$ determines the form of the current density through Eq.~\ref{A1}, we can guess the following decomposition of $\Vec{X}=\alpha \hat{e}+\beta \hat{\omega}+\gamma(\hat{e}\times\hat{\omega})$ with the unknowns $\alpha$, $\beta$, and $\gamma$. The mathematical steps for the calculation of $\alpha$, $\beta$, and $\gamma$ and the subsequent determination of current density runs exactly similar to the non-relativistic calculation done in Ref.~\cite{Dwibedi:2023akm}. Therefore, the main results will be written here, and a detailed calculation will be carried out in the appendix~\ref{ape1}. The total conductivity for the rotating HRG derived in appendix~\ref{ape1} is given by,
 \bea
 && \sigma^{ij}=\sigma^{0} \delta_{ij} +\sigma^{1}\epsilon_{ijk}\omega_k +\sigma^{2}\omega_i\omega_j,\nn\\
\text{with } &&\sigma^{n} =\sum\limits_{r} \frac{g_{r} q_{r}^2}{3T}\int \frac{d^{3}p}{(2\pi)^3}\frac{\tau_c\big(\frac{\tau_c}{\tau_{\Omega}}\big)^n} {1+\big(\frac{\tau_c}{\tau_{\Omega}}\big)^2}\times \frac{p^2}{E^{2}}f^{0}_{r}(1+af^{0}_{r})
\label{A11}
 \eea
 For the angular velocity in the z-direction i.e., $\Vec{\Om}=\Om \hat{k}$, the conductivity matrix has the following form,
\begin{equation}
  [\sigma]=\begin{pmatrix}
      \sigma^{0} & \sigma^{1} & 0\\
    
       -\sigma^{1} & \sigma^{0} & 0\\
      0 & 0 & \sigma^{0}+\sigma^{2}
  \end{pmatrix}~.
  \label{A13}
\end{equation}

\subsection{Electrical conductivity for HRG}\label{hrg}

 A quick glance at the matrix in Eq ~\ref{A13} led us to define the following conductivity components: parallel conductivity (parallel to angular velocity $\Vec{\Omega}$) $\sigma^{||}\equiv \sigma^{0}+\sigma^{2}$, perpendicular conductivity (perpendicular to angular velocity $\Vec{\Omega}$) $\sigma^{\perp}\equiv \sigma^{0}$ and Cross or Hall like conductivity $\sigma^{\times}\equiv \sigma^{1}$. Moreover, one can identify $\sigma_{||}$ with the conductivity in the absence of $\Omega$ i.e, $\sigma^{||}\equiv \sigma$.  

In Eq.~\ref{A11}, we have derived the conductivity tensor for the rotating hadron gas. We can rewrite this equation with two separate summations for the baryons and mesons, respectively, along with their spin degeneracy factors. The parallel conductivity  of the rotating HRG (or the HRG conductivity in the absence of $\Omega$) at $\mu=0$ is given by,
\be
\sigma^{||}_{\rm HRG}\equiv \sigma_{\rm HRG}  = \sum_{B} \frac{g_B q_{B}^2}{3T}\int \frac{d^{3}p}{(2\pi)^3}\tau_c\times \frac{p^2}{E^{2}}f^0(1-f^0)
+ \sum_{M} \frac{g_M q_{M}^2}{3T}\int \frac{d^{3}p}{(2\pi)^3}\tau_c\times \frac{p^2}{E^{2}}f^0(1+f^0)
\label{s_HRG}
\ee
where $g_H$ is spin degeneracy of hadrons with charges $q_H$, masses $m_H$ and energy $E=\sqrt{p^2+m_H^2}$. For $H=$ Mesons
and Baryons, equilibrium distribution function will be $f_0=\frac{1}{e^{E/T}- 1}$ and $f_0=\frac{1}{e^{E/T}+ 1}$ respectively.
Hadrons with a neutral electric charge will not participate in electrical conductivity. 

The relaxation time of any hadron can be written as
\be
\tau_c=1/(n_{HRG} v^H_{\rm av}\pi a^2)~,
\label{tc_HRG}
\ee 
where hard sphere cross-section $\pi a^2$ is considered for hadron, having
average velocity
\be 
v^H_{\rm av}=\int \frac{d^3p}{(2\pi)^3}\frac{p}{E} f_0\Big/\int \frac{d^3p}{(2\pi)^3} f_0~.\label{v_av}
\ee 
Each hadron will face the entire density of the system
\be
n_{HRG}=\sum_{B}g_B\int_{0}^{\infty}\frac{d^3p}{(2\pi)^3}\frac{1}{e^{E/T}+1} +\sum_{M} g_M\int_{0}^{\infty}\frac{d^3p}{(2\pi)^3}\frac{1}{e^{E/T}-1} 
\label{n_HRG}
\ee
where $g_B$ and $g_M$ are Baryon and Meson spin degeneracy factors, respectively.

The perpendicular electrical conductivity of the rotating HRG can be written as:
\be
\sigma^{\perp}_{HRG} =\sum_{B} \frac{g_B q_{B}^2}{3T}\int \frac{d^{3}p}{(2\pi)^3}\frac{\tau_c}{1+\big(\frac{\tau_c}{\tau_{\Omega}}\big)^2}\times \frac{p^2}{E^{2}}f^0(1-f^0)
+\sum_{M} \frac{g_M q_{M}^2}{3T}\int \frac{d^{3}p}{(2\pi)^3}\frac{\tau_c}{1+\big(\frac{\tau_c}{\tau_{\Omega}}\big)^2}\times \frac{p^2}{E^{2}}f^0(1+f^0)
\label{sO_HRG}
\ee
Similarly, the Hall electrical conductivity can be expressed as:
\be
\sigma^{\times}_{HRG} =\sum_{B} \frac{g_B q_{B}^2}{3T}\int \frac{d^{3}p}{(2\pi)^3}\frac{\tau_c\big(\frac{\tau_c}{\tau_{\Omega}}\big)}{1+\big(\frac{\tau_c}{\tau_{\Omega}}\big)^2}\times \frac{p^2}{E^{2}}f^0(1-f^0)
+\sum_{M} \frac{g_M q_{M}^2}{3T}\int \frac{d^{3}p}{(2\pi)^3}\frac{\tau_c\big(\frac{\tau_c}{\tau_{\Omega}}\big)}{1+\big(\frac{\tau_c}{\tau_{\Omega}}\big)^2}\times \frac{p^2}{E^{2}}f^0(1+f^0)
\label{sOH_HRG}
\ee
\subsection{Electrical conductivity for mass-less QGP}\label{mqgp}

 We can construct the conductivity tensor for the massless rotating QGP by substituting $p=E$ in Eq.~\ref{A11} and summing over all the light quarks with their corresponding degeneracies as
\begin{eqnarray}
 \sigma_{\rm QGP}^{n} &&=\sum_{f=u,d,s} \frac{g_f q_f^2 }{6\pi^{2}T}\frac{\tau_c\big(\frac{\tau_c}{\tau_{\Omega}}\big)^n}{1+\big(\frac{\tau_c}{\tau_{\Omega}}\big)^2}\int dE E^2f^{0}(1-f^{0})\nn\\
&&=\sum_{f=u,d,s} \frac{g_f q_f^2 }{6\pi^{2}T}\frac{\tau_c\big(\frac{\tau_c}{\tau_{\Omega}}\big)^n}{1+\big(\frac{\tau_c}{\tau_{\Omega}}\big)^2}T\frac{\partial}{\partial\mu}\int dE E^2f^{0}\nn\\
&&=\sum_{f=u,d,s} \frac{g_f q_f^2 }{6\pi^{2}T}\frac{\tau_c\big(\frac{\tau_c}{\tau_{\Omega}}\big)^n}{1+\big(\frac{\tau_c}{\tau_{\Omega}}\big)^2}T \frac{\partial}{\partial\mu}T^{3} \Gamma(3) f_{3}(A),\big( \text{ where } f_j(A)=\frac{1}{\Gamma(j)}\int_{0}^{\infty}\frac{x^{j-1}dx}{A^{-1}e^{x}+1}, (A\equiv e^{\mu/T})\big)\nn\\
&&=\sum_{f=u,d,s} \frac{g_f q_f^2 }{6\pi^{2}T}\frac{\tau_c\big(\frac{\tau_c}{\tau_{\Omega}}\big)^n}{1+\big(\frac{\tau_c}{\tau_{\Omega}}\big)^2}T\frac{\partial}{\partial\mu}T^3 \Gamma(3) f_{3}(A)\nn\\
&&=\sum_{f=u,d,s} \frac{g_f q_f^2 }{3\pi^{2}}\frac{\tau_c\big(\frac{\tau_c}{\tau_{\Omega}}\big)^n}{1+\big(\frac{\tau_c}{\tau_{\Omega}}\big)^2}T^2 f_{2}(A)
\label{A14}   
\end{eqnarray}

The electrical conductivity for massless QGP in the absence of rotation (or parallel conductivity of rotating QGP) is defined as,
\bea
\sigma_{\rm QGP}\equiv \sigma^{||}_{\rm QGP}= \sigma^{0}_{\rm QGP}+\sigma^{2}_{\rm QGP}&=& \frac{1}{3\pi^{2}}\Big(\sum_{f=u,d,s} g_f q_f^2\Big)\tau_c T^2 L_{2}
\nn\\
&=& \frac{8e^2}{3\pi^{2}}\tau_c T^2 L_{2}
\label{s_QGP}
\eea
where $g_f=$ spin degeneracy $\times$ color degeneracy $\times$ particle-anti-particle degeneracy$ = 2\times 3\times 2 = 12$
for any quark flavor $f=u, d, s$ with charges $q_u=+\frac{2}{3}e$, $q_d=-\frac{1}{3}e$, $q_s=-\frac{1}{3}e$. In natural unit
$e^2=4\pi/137$. Being electric charge neutral, gluons will not participate in electrical conductivity. At $\mu=0$, Fermi integral
function $f_n$ will convert to $L_n$: 
\be
L_j=\frac{1}{\Gamma(j)}\int_{0}^{\infty}\frac{x^{j-1}dx}{e^{x}+1}=f_{j}(1)~,
\ee
where we have chosen $a=-1$, since quarks are fermions.\\
Now, quarks will face the entire QGP density
\bea
n_{QGP}&=&g_q\int_{0}^{\infty}\frac{d^3p}{(2\pi)^3}\frac{1}{e^{p/T}+1} +g_g\int_{0}^{\infty}\frac{d^3p}{(2\pi)^3}\frac{1}{e^{p/T}-1} 
\nn\\
&=& \Big[g_q\Big(1-\frac{1}{2^{3-1}}\Big)+g_g\Big]\frac{\zeta(3)}{\pi^2}T^3~,
\nn\\
&=& \Big[\frac{3g_q}{4}+g_g\Big]\frac{\zeta(3)}{\pi^2}T^3~,
\label{n_QGP}
\eea 
where $g_q=2\times 3\times 2\times 3=36$ and $g_g=2\times 8=16$ are quark and gluon degeneracy factors respectively and
Reimann Zeta function $\zeta(3)=1.202$.

One can write perpendicular electrical conductivity $\sigma^{\perp}\equiv \sigma^{0}$ for rotating QGP from Eq.~\ref{A14} as:
\bea
\sigma^{\perp}_{\rm QGP}&=& \frac{1}{3\pi^{2}}\Big(\sum_{f=u,d,s} g_f q_f^2\Big)\frac{\tau_c}{1+\big(\frac{\tau_c}{\tau_{\Omega}}\big)^2} T^2 L_{2}
\nn\\
&=& \frac{8e^2}{3\pi^{2}}\frac{\tau_c}{1+\big(\frac{\tau_c}{\tau_{\Omega}}\big)^2} T^2 L_{2}
\label{sO_QGP}
\eea
Similarly, from Eq.~\ref{A14} the Hall electrical conductivity $\sigma^{\times}\equiv \sigma^{1}$ can be expressed as,
\bea
\sigma^{\times}_{\rm QGP}&=& \frac{1}{3\pi^{2}}\Big(\sum_{f=u,d,s} g_f q_f^2\Big)\frac{\tau_c\big(\frac{\tau_c}{\tau_{\Omega}}\big)}{1+\big(\frac{\tau_c}{\tau_{\Omega}}\big)^2} T^2 L_{2}
\nn\\
&=& \frac{8e^2}{3\pi^{2}}\frac{\tau_c\big(\frac{\tau_c}{\tau_{\Omega}}\big)}{1+\big(\frac{\tau_c}{\tau_{\Omega}}\big)^2} T^2 L_{2}
\label{sOH_QGP}
\eea

\section{Results and Discussion}
\label{Sec:Results_Discussion}
For numerical evaluation of electrical conductivities for a rotating QGP, we have employed the formulas put down in \textbf{Sec.~\ref{mqgp}}. Similarly, for quantitative estimation of electrical conductivities for the rotating hadron gas, we use the Ideal Hadron Resonance Gas (IHRG) model established in \textbf{Sec.~\ref{hrg}}, which encompasses all the non-interacting hadrons and their resonance particles up to a mass of 2.6 GeV as listed in Ref.~\cite{ParticleDataGroup:2008zun}.
\begin{figure}[htp!]
	\centering 
	\hspace{0.cm} 
 	\includegraphics[width=0.6\textwidth]{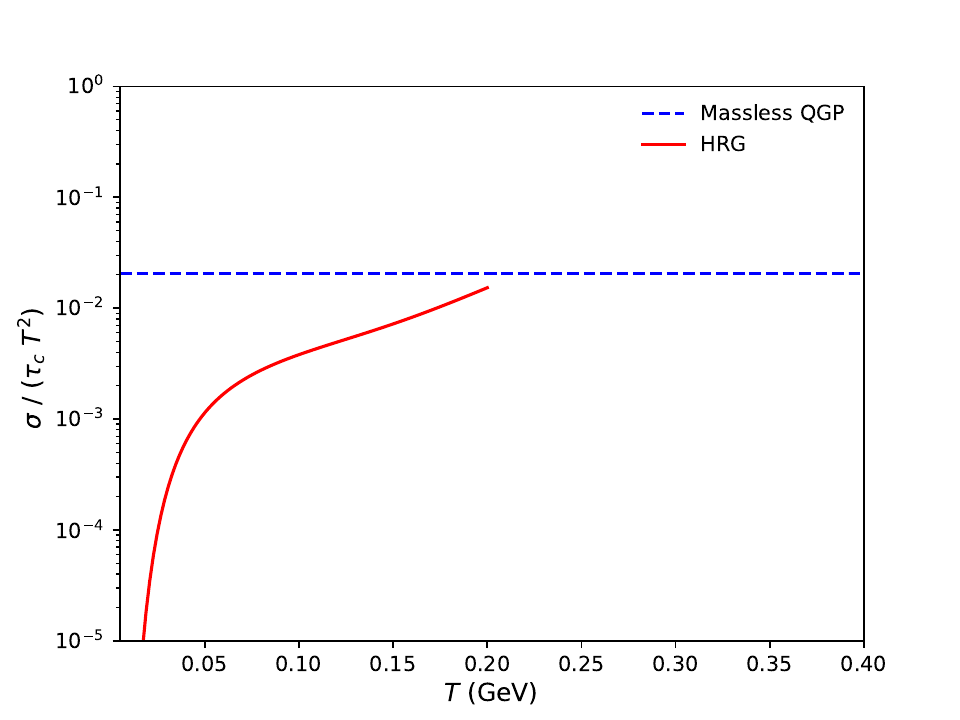}
	\caption{(Color online) Electrical conductivity $\sigma$ normalized by $\tau_c T^2$ as a function of $T$ for massless QGP and HRG.}
	\label{figure1}
\end{figure}
\begin{figure}[htp!]
	\centering 
	\hspace{0.cm} 
	\includegraphics[width=0.6\textwidth]{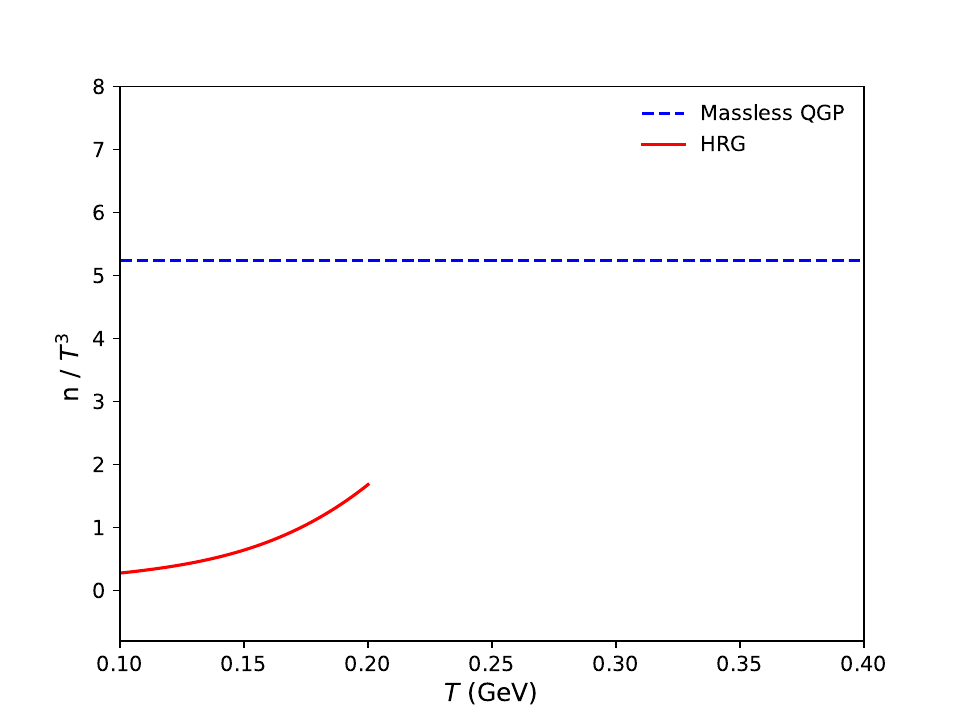}
	\caption{(Color online) Normalized number density $n/(T^3)$ vs $T$ for massless QGP and HRG.}
	\label{figure2}
\end{figure}

In Fig.~\ref{figure1}, we have portrayed the normalized electrical conductivity $\sigma/\tau_{c} T^{2}$ with respect to temperature $T$. We take the help of Eq.~\ref{s_QGP} to get the expression of $\sigma_{\rm QGP}$ for massless QGP, which can be seen to be directly proportional to $\tau_{c}T^{2}$. Accordingly, we have obtained a horizontal line corresponding to $\sigma/\tau_{c} T^{2}=0.02$. For the hadronic temperature regime, we have presented the variation of scaled conductivity $\sigma/\tau_{c} T^{2}$ by resorting to the HRG model Eq.~\ref{s_HRG} at zero baryon chemical potential. For simplicity, we have assumed constant $\tau_{c}$ for all the hadrons to obtain the pattern of scaled $\sigma_{\rm HRG}$ in Fig.~\ref{figure1}. The plot (red solid line) displays a sharp rise at low temperatures and eventually flattens as the temperature increases. The conductivity for the hadron gas obtained from the HRG model stays below the massless QGP (blue dashed line). The pattern is quite similar to the normalized thermodynamical quantities like pressure, energy density, etc, whose HRG estimations always remain below their massless QGP or Stefan-Boltzmann (SB) limits.
\begin{figure}[htp!]
	\centering 
	\hspace{0.cm} 
	\includegraphics[width=0.6\textwidth]{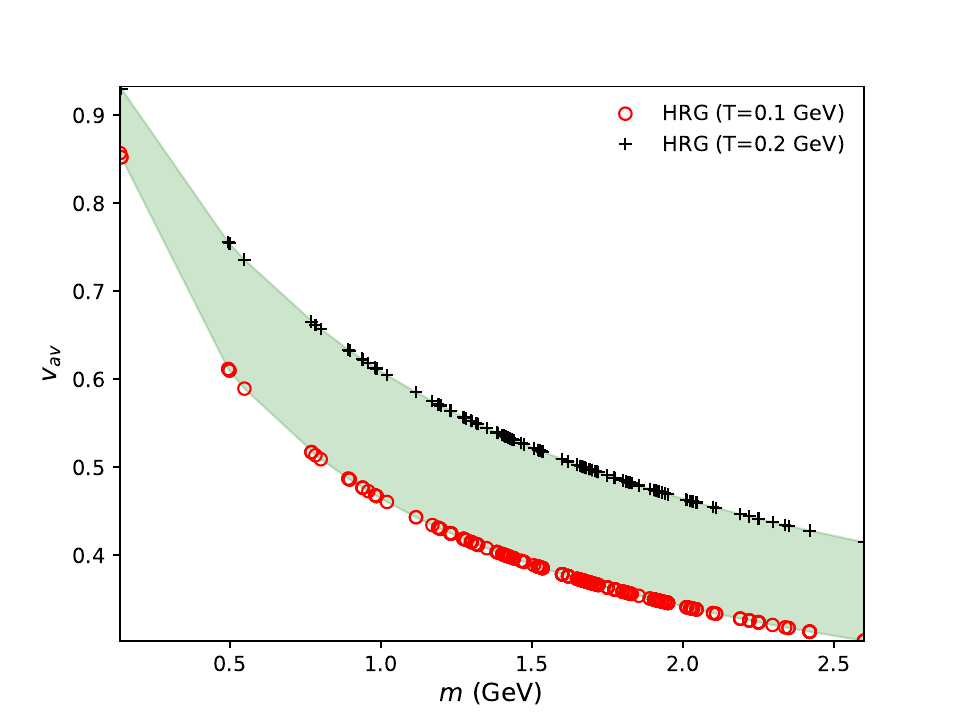}
	\caption{(Color online)  Average velocity ($v_{\text{av}}$) as a function of hadron mass (m) for all hadrons at two different temperatures: $T$ = 0.1 GeV and T = 0.2 GeV.}
	\label{figure3}
\end{figure}

In Fig.~\ref{figure2}, we have compared the numerical magnitudes of normalized number density $\frac{n}{T^{3}}$ for massless QGP (red solid line) and hadron gas (blue dashed line). We use Eq.~\ref{n_QGP} and Eq.~\ref{n_HRG} for massless QGP and hadron gas for the determination of the magnitude of $\frac{n}{T^{3}}$. For QGP, owing to the relation $n \propto T ^{3}$, we get a horizontal line at $\frac{n}{T^{3}}\approx 5.23$. In contrast, in the hadron gas, our model calculation produces a monotonically increasing $n$ or $n/T ^{3}$ with respect to $T$.  Again, we observe that, like conductivity, the number density value for hadron gas estimated from the HRG model stays below the massless QGP limit. 

In Fig.~\ref{figure3}, we have displayed the variation of the average velocity of different hadrons in the hadron gas with respect to their masses up to 2.6 GeV at two different temperatures: $T=0.1$ GeV and $T=0.2$ GeV. We have obtained the numerical values of $v_{\text{av}}$ from Eq.~\ref{v_av} at zero baryonic chemical potential. The result shows a decrease in average velocity $v_{\text{av}}$ of all the hadrons with respect to their masses. The lighter hadrons have high velocities compared to the heavier ones. An increase in temperature makes hadrons move with an increased velocity as the thermal energy increases with temperature. Moreover, we have also created a band to depict the range of velocity of all hadrons between $T=0.1$ GeV and $T=0.2$ GeV.

\begin{figure}[htp!]
	\centering 
	\hspace{0.cm} 
	\includegraphics[width=0.6\textwidth]{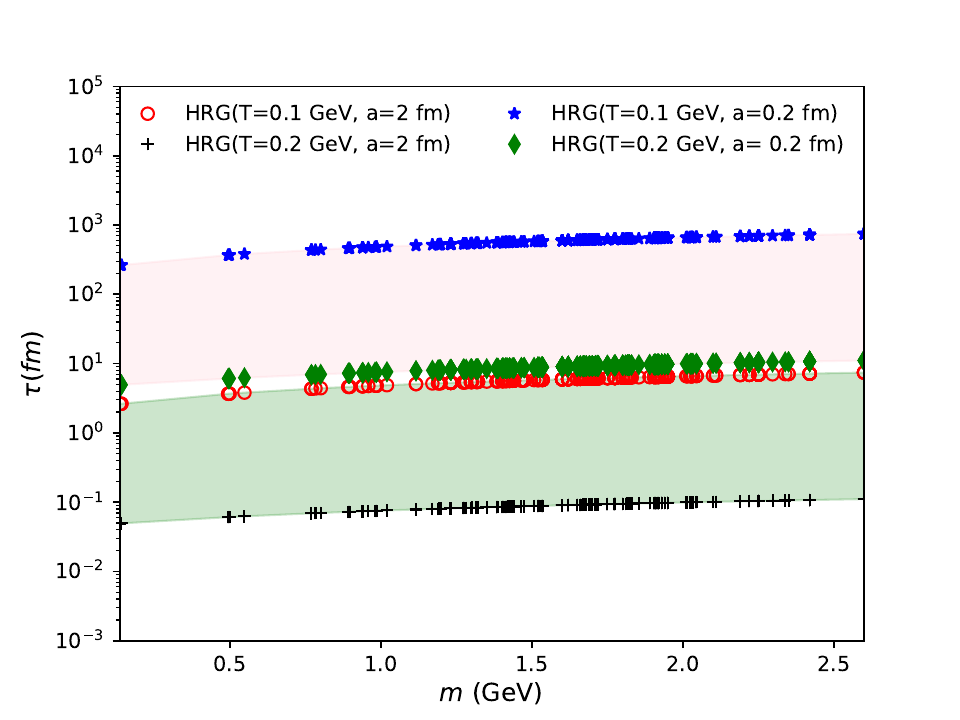}
	\caption{(Color online) Relaxation Time $\tau_{c}$ vs (m) for various hadrons at two different temperatures: $T$ = 0.1 GeV and T = 0.2 GeV. Two different values of $a$ ($0.2$ fm and $2$ fm) are considered for each temperature.}
	\label{figure4}
\end{figure}
\begin{figure}[htp!]
	\centering 
	\hspace{0.cm} 
	\includegraphics[width=0.6\textwidth]{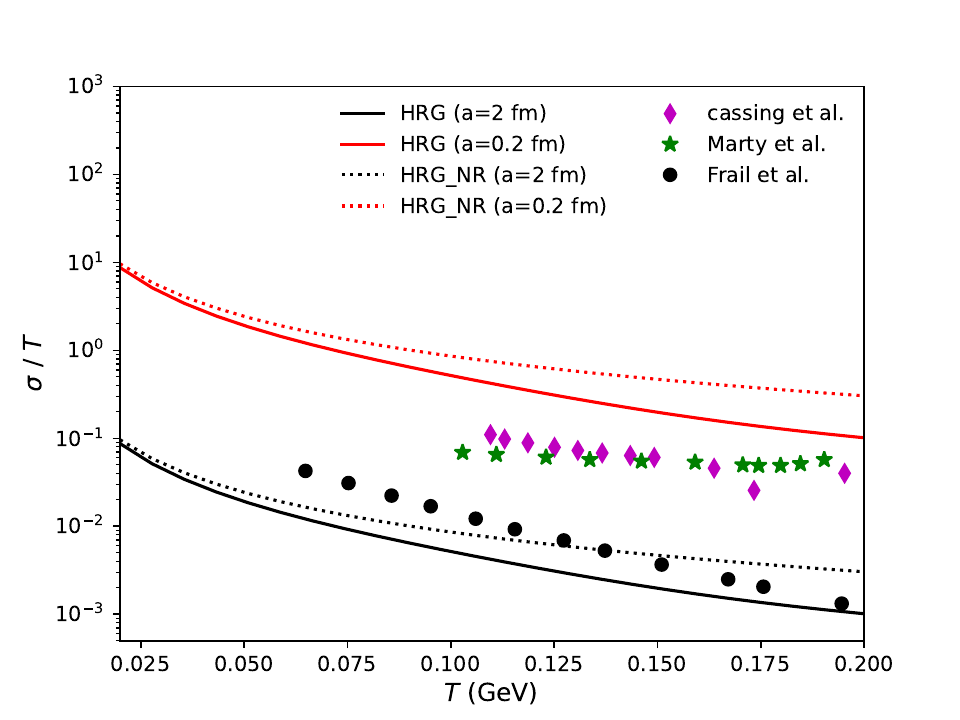}
	\caption{(Color online)  Electrical conductivity $\sigma/ T$ vs $T$ for HRG with temperature-dependent $\tau_c(T)$ at a = 0.2 and 2 fm and its comparison with the models of Cassing et al.~\cite{Cassing:2013iz}, Marty et al.~\cite{Marty:2013ita}, Frail et al.~\cite{Fernandez-Fraile:2005bew}. Along with the relativistic estimations (solid lines), a non-relativistic estimations (dotted lines) are also added to show the numerical contribution of relativistic correction.}
	\label{figure5}
\end{figure}

In Fig.~\ref{figure4}, we have illustrated the change of relaxation or collision time of different hadrons with respect to their masses for two different temperature values: $T=0.1$ GeV and $T=0.2$ GeV. For the evaluation of collision time, we have relied on the expression of the hard-sphere scattering model of the collision set down in Eq.~\ref{tc_HRG}. For each $T$ value, we take two different scattering lengths $a$ to display the variation of $\tau_{c}$ with respect to both $T$ and $a$. For a fixed $a$ the relaxation time $\tau_{c}(T)$ decreases because of the increase of $v_{\text{av}}(T)$ and $n(T)$ by following the Eq.~\ref{tc_HRG}. Similarly, for a given $T$, the relaxation time increases with $a$ since $\tau_{c}\propto \frac{1}{a^{2}}$. Here, we have chosen $a=0.2$ fm and $a=2$ fm, whose reason will be clear in Fig.~\ref{figure5}. 

Next, we have aimed to compile earlier estimated data of $\sigma/T$ within the hadronic temperature domain, where few selective estimated results \cite{Cassing:2013iz,Marty:2013ita,Fernandez-Fraile:2005bew} are shown in Fig.~\ref{figure5}. The order of magnitude for $\sigma/T$, obtained by Cassing et al. \cite{Cassing:2013iz} (Diamonds), Marty et al. ~\cite{Marty:2013ita} (Stars), Fraile et al. ~\cite{Fernandez-Fraile:2005bew} (solid circles) are within the range $0.001$ to $0.1$. A long list of references~\cite{Cassing:2013iz,Marty:2013ita,Fernandez-Fraile:2005bew,PhysRevD.83.034504,PhysRevLett.99.022002,PhysRevLett.105.132001,Burnier2012,GUPTA200457,Brandt2013,PhysRevLett.111.172001,PhysRevC.90.044903,PhysRevD.90.114009,PhysRevD.90.094014,PhysRevD.89.106008,PhysRevC.90.025204} can be found for microscopic estimation of $\sigma/T$, whose order of magnitude will be located within $0.001$ to $0.1$ for hadronic temperature domain and $0.002$ to $0.3$ within quark temperature domain. Now, it can be seen that all the data obtained from earlier works within the hadronic temperature domain can be covered by altering $a$ from $0.2$ to $2$ fm. For this reason, the same range of $a$ has been considered in previous Fig.~\ref{figure4}. We have also added a non-relativistic version of HRG estimations (dotted lines) along with our relativistic estimations (solid lines).
Since the present framework may be considered as a relativistic extension of earlier Ref.~\cite{Dwibedi:2023akm}, which was done in a non-relativistic framework,
so to see their difference, we have quickly gone through an HRG model in a non-relativistic-framework. Its quick formalism part is added in appendix \ref{ape2}.
\begin{figure}[htp!]
	\centering 
	\hspace{0.cm} 
	\includegraphics[width=0.6\textwidth]{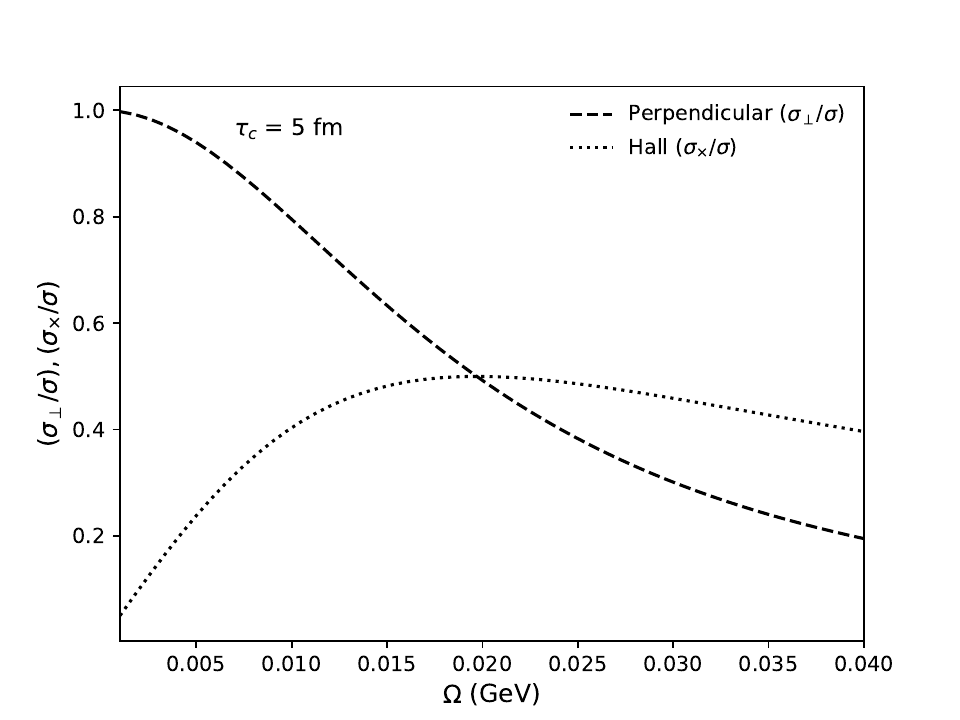}
	\caption{Perpendicular and Hall Electrical conductivity $\sigma_{\perp}/(\sigma)$, $\sigma_{\times}/(\sigma)$ vs $\Omega$ for HRG at $\tau_c=5$ fm and $T=0.150$ GeV.}
	\label{figure6}
\end{figure} 

Upto Fig.~\ref{figure5}, we have gone through the estimations of $\sigma/T$ in the absence of rotation. In presence of rotation the isotropic nature of $\sigma/T$ converts into anistropic nature of $\sigma/T$, having multi components- $\sigma^{||}$, $\sigma^{\perp}$ and $\sigma^{\times}$. Interestingly, $\sigma^{||}$ in the presence of rotation is the same as $\sigma$, which is the isotropic conductivity in the absence of $\Omega$. The expression of $\sigma (T)$ given in Eq.~\ref{s_HRG} has two components: thermodynamical phase space and relaxation time $\tau_{c}(T)$. The former component has a non-tunable temperature profile, while the latter component can be tunable by tuning the magnitude of scattering cross-section through $a$. We have used hard sphere scattering cross-section relation for the expression of $\tau_{c}(T)$, given in Eq.~\ref{tc_HRG}. The temperature dependance of $\tau_{c}$ is mainly determined by $n(T)$ and $v_{\rm av}(T)$ which are displayed in the earlier Figs.~\ref{figure2} and \ref{figure3}. After calibrating our results without rotation with earlier estimations, we will now proceed to apply them to rotating hadronic matter.

In Fig.~\ref{figure6}, we have depicted the variation of perpendicular and hall conductivities $\sigma^{\perp,\times}$ normalized by the parallel conductivity $\sigma^{||}=\sigma$ as a function of $\Omega$ for a fixed $T=0.150$ GeV. We have employed Eq.~\ref{sO_HRG},\ref{sOH_HRG} and \ref{s_HRG} for the numerical estimation of $\sigma^{\perp}$, $\sigma^{\times}$ and $\sigma$ respectively for rotating HRG. In relation to Eq.~\ref{tc_HRG} and Fig.~\ref{figure4}, it is apparent that the relaxation time is a function of $a$, $T$, and $m$, i.e., $\tau_{c}=\tau_{c}(a,m,T)$. So, for a specific hadron in the system at a given temperature, it depends on the effective hard sphere scattering length. In the beginning, let us consider a constant $\tau_{c}$ for estimating $\sigma^{||,\perp,\times}$ instead of the actual $\tau_{c}(a,m,T)$. By doing this, we can visualize only the thermodynamical phase space part of $\sigma^{||,\perp,\times}(T,\Omega)$. In this plot, we choose a value of $\tau_{c}=5$ fm (25 GeV$^{-1}$), which falls in the band of $\tau_{c}$ obtained in Fig.~\ref{figure4} for $a=2$ fm. In HICs, the average value of the local vorticity can be taken as a measure of the global vorticity or angular velocity of the system. The average vorticity for HIC has been calculated from various models~\cite{XuGuangHuang:2020dtn,PhysRevC.95.054915,XuGuangHuang2016,Jiang2016}. Inspired by these studies, we choose the scale of the $\Omega$ axis from $0-0.04$ GeV. $\sigma^{\perp}$ (or $\sigma^{0}$), the perpendicular conductivity of the rotating HRG corresponds to the current in the direction of the applied electric field in the $XY$ plane. In the limit of $\Omega\xrightarrow{}0$, $\sigma^{\perp}$ reduces to conductivity $\sigma$. In our plot, this feature can be seen where $\frac{\sigma^{\perp}}{\sigma} \xrightarrow{}1$ as one approach to $\Omega\xrightarrow{}0$. Similarly, we can see the matrix in Eqn.~\ref{A13} that $\sigma^{1}\equiv \sigma^{\times}$ drives the electric current in the XY-plane of the rotating hadron gas; it drives current in the $X$-direction if the electric field is in the $Y$-direction and vice-versa. In the limit  $\Omega\xrightarrow{}0$, $\sigma^{\times}$ vanishes. The Hall conductivity $\sigma^{\times}$ shows interesting characteristics, it first increases with $\Omega$ to hit a peak where $\tau_{\Omega}\equiv\frac{1}{2\Omega}$ approaches $\tau_{c}$ and then decreases with further increase in $\Omega$. From Fig.\ref{figure6} one can notice that, for slowly rotating HRG, i.e., at low $\Omega$, the $\sigma^{\perp}$ dominates over $\sigma^{\times}$ whereas, for a fastly rotating HRG, i.e., at high $\Omega$ we see that $\sigma^{\perp}< \sigma^{\times}$. Noticeably, the magnitude of $\sigma^{\perp,\times}/\sigma$ in the figure lies below one, i.e., $\sigma^{\perp,\times}/\sigma \leq 1$.  This property can be understood by recognizing three different time scales associated with the rotating hadronic gas: $\tau_{c}\equiv \tau^{||}_{c}$, $\tau^{\perp}_{c}$ and $\tau^{\times}_{c}$. The effective relaxation times $\tau^{\perp}_{c}$ and $\tau^{\times}_{c}$  occurs in the mathematical expression of $\sigma^{\perp}$ and $\sigma^{\times}$ are given below,
\bea
&& \tau^{\perp}_{c}= \frac{\tau_{c}}{1+ \big(\frac{\tau_{c}}{\tau_{\Omega}}\big)^{2}},\nn\\
&& \tau^{\times}_{c}= \frac{\tau_{c}\big(\frac{\tau_{c}}{\tau_{\Omega}}\big)}{1+ \big(\frac{\tau_{c}}{\tau_{\Omega}}\big)^{2}}.\nn
\eea
A glance at the above time scales suggests that $\tau^{\perp}_{c}<\tau_{c}$ and $\tau^{\times}_{c}<\tau_{c}$ which determines the ordering $\sigma^{\perp}<\sigma$ and $\sigma^{\times}<\sigma$.
\begin{figure}[htp!]
	\centering 
	\hspace{0.cm} 
	\includegraphics[width=0.6\textwidth]{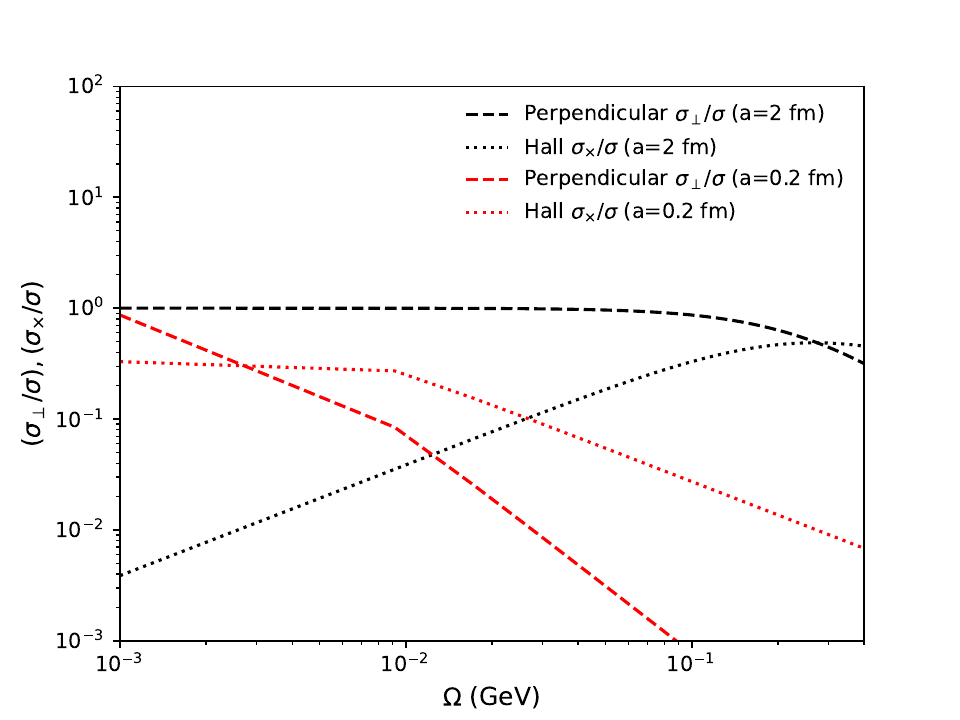}
	\caption{(Color online) Perpendicular and Hall Electrical conductivity ($\sigma_{\perp}/\sigma,\sigma_{\times}/\sigma$) vs $\Omega$ for HRG at T=0.150 ${\rm~ GeV}$.}
	\label{figure7}
\end{figure}

 We have delineated the variation of normalized conductivity $\frac{\sigma^{\perp,\times}}{\sigma}$ in relation to the angular velocity $\Omega$ of the hadronic medium at $T=0.150$ GeV in Fig.~\ref{figure7}. In contrast to Fig.~\ref{figure6}, where we have interpreted the alternation of $\frac{\sigma^{\perp,\times}}{\sigma}$ graphically for a fixed value of $\tau_{c}=5$ fm, we take here the individual $\tau_{c}$ of the rotating hadrons by using Eq.~\ref{tc_HRG}. For fixed $T=0.150$ GeV and take two different values $a=(0.2~ {\rm fm},2~ {\rm fm})$ we have calculated $\tau_{c}=\tau_{c}(a,m,T)$ for different hadron resonances with mass $m$. The different values of $a$ chosen here are the tuned scattering lengths obtained by the calibration done in Fig.~\ref{figure5}. For the value of $a=2$ fm, we see that for a rotating HRG with $\Omega$ in the range $0.001$ GeV to $0.02$ GeV, the $\sigma^{\perp}$ is almost equal to $\sigma$ and $\sigma^{\times}$ is negligible. This suggests an almost isotropic HRG with the scattering length $a=2$ fm. Nevertheless, for $a=0.2$ fm, we observe that in the range $\Omega= 0.01-0.02$ GeV there is a significant magnitude of $\sigma^{\times}$ which is around $10$ to $15$ \% of $\sigma$. Also, in the same range of $\Omega$, one can notice a significant suppression of the $\sigma^{\perp}$ with respect to $\sigma$, which is around $90\%$. This suggests a highly anisotropic HRG with a hugely suppressed perpendicular conductivity $\sigma^{\perp}$ along with a large magnitude of hall conductivity $\sigma^{\times}$. 
\begin{figure}[htp!]
	\centering 
	\hspace{0.cm} 
	\includegraphics[width=0.6\textwidth]{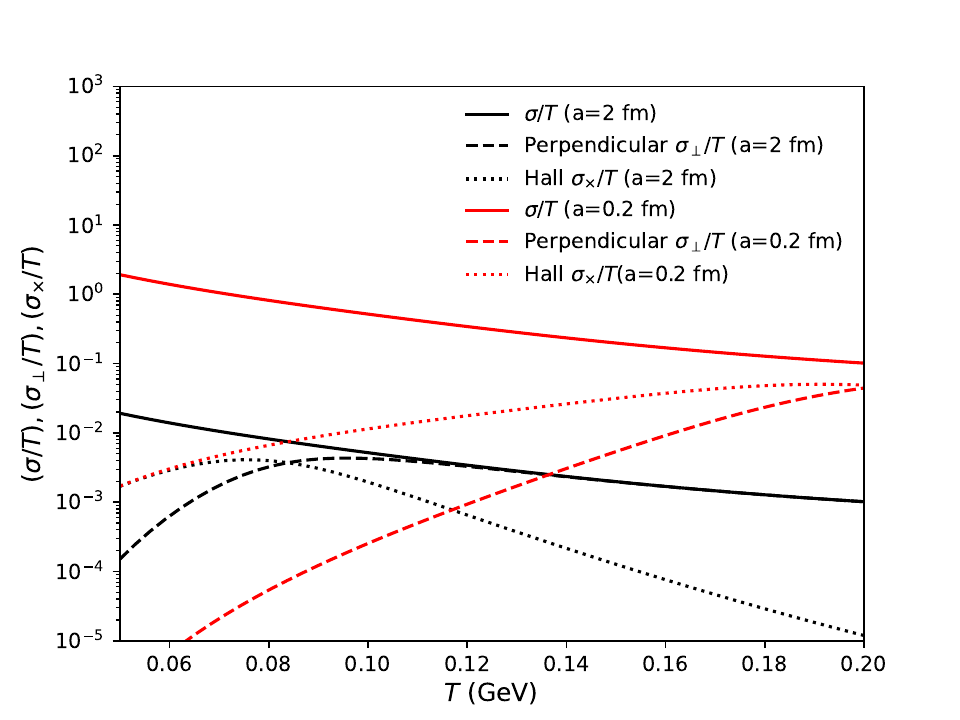}
	\caption{(Color online) Parallel, perpendicular, and Hall electrical conductivity ($\sigma/T$, $\sigma_{\perp}/T$, $\sigma_{\times}/T$) as functions of temperature ($T$) for HRG with $\tau_{c}(T)$ and $\tau_{\Omega} = 6$ fm.}
	\label{figure8}
\end{figure}
In Fig.~\ref{figure8}, we have displayed the temperature dependence of the scaled conductivities $\sigma/T$ for the rotating HRG at an angular speed $\Omega=0.016$ GeV ($\tau_{\Omega}\approx6$ fm). Similar to the previous figure, we take two different values of $a~(0.2~ {\rm fm},2~ {\rm fm})$ for the determination of $\tau_{c}(a,M,T)$ from  Eq.~\ref{tc_HRG}. The plot represents a strong variation of both perpendicular $\sigma^{\perp}$ and hall $\sigma^{\times}$ conductivities in relation to temperature $T$. For a rotating HRG with $a=2$ fm, the $\sigma^{\perp}$ almost merges with $\sigma$ for temperature $T\geq 0.12$ GeV. This implies an almost isotropic rotating HRG with $\sigma^{\perp}\approx \sigma$. Nevertheless, one can define a region of temperature $T=0.05-0.14$ GeV where the system still has a significant magnitude of $\sigma^{\times}$ with $\sigma/\sigma^{\times}\sim 10$ or less. On the other hand, in a rotating HRG with $a=0.2$ fm, there is a remarkable suppression of $\sigma^{\perp}$ with respect to $\sigma$ in the $T$ range $0.10-0.20$ GeV. The magnitude of the suppression is around $90\%$ at $T=0.12$ GeV. These observations suggest strongly anisotropic rotating HRG with $a=0.2$ fm. There is also a significant magnitude of hall conductivity $\sigma^{\times}$ for $T$ more than $0.14$ GeV, i.e., $\sigma/\sigma^{\times}\leq 10$ for $T\geq 0.14$ GeV. 

In the end, we can find a succinct qualitative message from our detailed quantitative investigations. It says that
Coriolis force can have a noticeable impact on creating an anisotropic conductivity tensor at a finite rotation of the HRG system, as done by the Lorentz force at a finite magnetic field. However, this noticeable impact will quantitatively depend on the hadronic scattering strengths, quantified by $a$ in our study. We have shown the noticeable and non-noticeable impact for $a=0.2$ fm and $a=2$ fm, respectively.
Another debatable point is that how fast or slow the angular velocity will decay with time~\cite{Jiang2016,XuGuangHuang2016,XuGuangHuang:2020dtn} so that hadronic matter will face it. In this context,
a possibility of noticeable impact for any values within the range $\Omega= 0.001-0.02$ GeV can be found.\\

\section{Summary}
\label{Sec:Summary}
In this work, we have made an effort to visualize the effect of Coriolis force on electrical conductivity in the Hadron Resonance Gas Model. The coefficient of proportionality between electrical current density and electrical field is known as electrical conductivity. Using the Boltzmann transport equation and the relaxation time approximation, we have calculated their microscopic expressions within the framework of kinetic theory based on their macroscopic formulations. The Coriolis force in rotating frames leads to similar anisotropy in the electrical conductivity tensor ($\sigma$) as the Lorentz force introduced in the presence of a magnetic field.   
The generated anisotropy is categorized into three parts: the Hall, perpendicular, and parallel components. The parallel component of electrical conductivity remains unaffected as it is independent of the relaxation time of the medium. However, we observe the variations in the scaled electrical conductivity of the perpendicular and Hall components with temperature in the presence of Coriolis force.
In the absence of rotation, the scaled electrical conductivity (scaled with $\tau_c T^2$) increases with temperature but stays below the results obtained for massless QGP or Stefan-Boltzmann (SB) limits similar to the normalized thermodynamical quantities like pressure, energy density, etc. The average velocity of particles decreases monotonically with mass, which leads to an increase in particle relaxation time as a function of their respective masses. We estimate the relaxation time using the hard-sphere scattering model. 
We observed that the earlier estimations of ($\sigma / T$) could be tuned by varying the scattering lengths $a$ from 0.2 fm to 2 fm.  
A monotonic decreasing trend is observed in electrical conductivity with temperature in the absence of rotation.
The presence of the Coriolis force induces an anisotropic nature in electrical conductivity. 
We observed that as the angular velocity ($\Omega$) of the rotating hadron gas system increases, the perpendicular component of electrical conductivity decreases. As the rotation speed ($\Omega$) approaches zero, the perpendicular component converges to the overall conductivity $\sigma$. On the contrary, the Hall component vanishes towards small $\Omega$. The Hall component shows interesting behavior; initially, it increases with $\Omega$, reaching a peak when the characteristic rotation time ($\tau_{\Omega}$) becomes comparable to the relaxation time $\tau_{c}$ and then decreases with further increase in $\Omega$. Therefore, we can conclude that in a slowly rotating HRG, with low $\Omega$, $\sigma^{\perp}$ dominates over the Hall conductivity $\sigma^{\times}$, whereas, for a fastly rotating (large $\Omega$) HRG $\sigma^{\times} > \sigma^{\perp}$. Even more interestingly, we observe that this flip from $\sigma^{\perp}$ dominance to $\sigma^{\times}$ dominance occurs at relatively smaller angular velocities for systems with smaller scattering lengths. In the end, we estimate the variation of electrical conductivity components with temperature at a fixed angular speed. We observed a strong variation in both perpendicular and Hall components with temperature. With the chosen angular speed $\Omega=0.016$ GeV ($\tau_{\Omega}\approx6$ fm) the $\sigma^{\perp}$ almost merges with $\sigma$ above for temperature $T\approx 0.12$ GeV for $a=2$ fm. This implies an almost isotropic rotating HRG with $\sigma^{\perp}\approx \sigma$. On the other hand, for $a=0.2$ fm, suppression of $\sigma^{\perp}$ with respect to $\sigma$ is notably strong up to $T\approx0.20$ GeV, which suggests strongly anisotropic rotating HRG with $a=0.2$ fm. 

\section{Acknowledgement}
NP and AD gratefully acknowledge the Ministry of Education (MoE), Govt. of India. The authors extend their thanks to Snigdha Ghosh for sharing valuable materials and insights on HRG model calculations.
\section{appendix}
\subsection{Calculation of current density}\label{ape1}
The current density $J_{i}$ can be readily calculated from Eq.~\ref{A1} after the determination of $\delta f$. To calculate $\delta f$ let us substitute the ansatz $\delta f=-\Vec{p}\cdot\vec{X} \big( \frac{\partial f^0}{\partial E} \big)$ in the Eq.~(\ref{R12}), 
\begin{flalign*}
    \frac{\partial f^0}{\partial  E}\frac{\Vec{p}}{p_{0}}\cdot(q\Vec{\Tilde{E}})+2\Vec{p}\cdot\left[\Vec{\Omega}\times\frac{\partial\delta f}{\partial\Vec{p}}\right]&=\frac{-\delta f}{\tau_{c}}\\
    \implies \frac{\partial f^0}{\partial E}\frac{\Vec{p}}{E}\cdot(q\Vec{\Tilde{E}}) +2\Vec{p}\cdot\left[\Vec{\Omega}\times\frac{\partial(-\Vec{p}\cdot\Vec{X}\frac{\partial f^0}{\partial E})} {\partial\Vec{p}}\right] &=\frac{\Vec{p}\cdot\Vec{X}}{\tau_{c}}\frac{\partial f^0}{\partial E},   
\end{flalign*}
which, after simplification, becomes,
\begin{equation}
    \left[\frac{q\Tilde{\Vec{E}}}{E}+2(\Vec{X}\times\Vec{\Omega})\right]\cdot\Vec{p}\frac{\partial f^0}{\partial E}=\frac{\Vec{X}}{\tau_{c}} \cdot\left(\Vec{p}\frac{\partial f^0}{\partial E}\right),
    \label{A4}
\end{equation}

since $\Vec{p}$ is arbitrary in Eq.~\ref{A4}, the following relation is valid for $\Vec{X}$, 
\begin{equation}
  \left[\frac{q\Tilde{\Vec{E}}}{E}+2(\Vec{X}\times\Vec{\Omega})\right]=\frac{\Vec{X}}{\tau_{c}}~.
  \label{A5}
\end{equation}

Substituting the result, $\Vec{X}\times\Vec{\Omega}=-\gamma\Omega\hat{e}+\gamma\Omega(\hat{\omega}\cdot\hat{e})\hat{\omega}+\alpha\Omega(\hat{e}\times\hat{\omega})$, in Eq.~\ref{A5}, we have,
\begin{eqnarray}
    \Big( \frac{q\Vec{\Tilde{E}}}{E}-2\Omega\gamma \Big) \hat{e}+2\gamma \Omega (\hat{\omega}\cdot\hat{e}) \hat{\omega} +2 \alpha \Omega (\hat{e} \times \hat{\omega}) =\frac{\alpha}{\tau_{c}}\hat{e}+\frac{\beta}{\tau_{c}}\hat{\omega}+\frac{\gamma}{\tau_{c}}(\hat{e}\times\hat{\omega})~.
    \label{A6}
\end{eqnarray}
By equating the coefficients of the linearly independent basis vectors in Eq.~\ref{A6}, we get, 
\begin{align*}
 	\frac{q\tilde{E}}{E}-\frac{\gamma}{\tau_{\Omega}} &= \frac{\alpha}{\tau_c}, &
 	\frac{\gamma}{\tau_{\Omega}}(\hat{\omega}\cdot\hat{e}) &= \frac{\beta}{\tau_c}, &
 	\frac{\alpha}{\tau_{\Omega}} &=\frac{\gamma}{\tau_c},
\end{align*}
where $\tau_{\Omega}\equiv \frac{1}{2\Omega}$.
Simplifying the above three equations, we get the following values for $\alpha,\beta,\text{ and }\gamma$
\begin{align*}
    \alpha &= \frac{\tau_c\big(\frac{q\tilde{E}}{E}\big)} {1+\big(\frac{\tau_c}{\tau_{\Omega}} \big)^2}, & 
 	\gamma &=\frac{\tau_c \big(\frac{\tau_c}{\tau_{\Omega}}\big) \big(\frac{q\tilde{E}}{E} \big)} {1+\big(\frac{\tau_c}{\tau_{\Omega}}\big)^2}, &
 	\beta &=\frac{\tau_c \big(\frac{\tau_c}{\tau_{\Omega}}\big)^2 (\hat{\omega}\cdot\hat{e}) \big(\frac{q\tilde{E}}{E}\big)}{1+\big(\frac{\tau_c}{\tau_{\Omega}}\big)^2}.
\end{align*}
By substituting the expressions of $ \alpha,\beta\text{, and }\gamma$, we get the explicit form of the $\delta f$ as, 

\begin{eqnarray}
    \delta f &=& -\vec{p}\cdot\vec{{X}} \frac{\partial f^0}{\partial E}\nn\\
	&=&-\frac{\partial f^0}{\partial E} \vec{p} \cdot (\alpha\hat{e} +\beta\hat{\omega}+ \gamma(\hat{e}\times \hat{\omega}))\nn\\
	&=&-\frac{1}{E}\frac{\partial f^0}{\partial E}\Bigg[\frac{q\tau_c\tilde{E}}{1+\big(\frac{\tau_c}{\tau_{\Omega}}\big)^2}\hat{e}\cdot\vec{p}+\frac{\tau_c \big(\frac{\tau_c}{\tau_{\Omega}}\big)^2 (\hat{\omega}\cdot\hat{e})q\tilde{E}}{1+\big(\frac{\tau_c}{\tau_{\Omega}}\big)^2} \hat{\omega}\cdot\vec{p}+\frac{\tau_c \big(\frac{\tau_c}{\tau_{\Omega}}\big) q \tilde{E}} {1+\big(\frac{\tau_c}{\tau_{\Omega}}\big)^2} (\hat{e} \times \hat{\omega})\cdot\vec{p}\Bigg]\nn\\
	&=&-\frac{1}{E}\frac{\partial f^0}{\partial E}\bigg(\frac{q \tau_c}{1+\big(\frac{\tau_c}{\tau_{\Omega}}\big)^2} \bigg) \Bigg[\vec{\tilde{E}}\cdot\vec{p}+\Big(\frac{\tau_c}{\tau_{\Omega}} \Big)^2 (\hat{\omega}\cdot\vec{\tilde{E}}) (\hat{\omega}\cdot \vec{p})+ \Big(\frac{\tau_c}{\tau_{\Omega}}\Big) (\vec{\tilde{E}} \times \hat{\omega})\cdot\vec{p} \Bigg]\nn\\
	&=&-\frac{1}{E}\frac{\partial f^0}{\partial E} \bigg( \frac{q\tau_c}{1+\big(\frac{\tau_c} {\tau_{\Omega}}\big)^2} \bigg) \bigg[ \tilde{E}_j p^{j}+\Big( \frac{\tau_c}{\tau_{\Omega}} \Big)^2 \omega_j \tilde{E}_j \omega_q p^q +\Big( \frac{\tau_c}{\tau_{\Omega}}\Big) \epsilon_{qjk}\tilde{E}_j\omega_k p^q\bigg]\nn\\
	&=& -\frac{q}{E}\frac{\partial f^0}{\partial E} \bigg(\frac{\tau_c}{1+ \big( \frac{\tau_c} {\tau_{\Omega}} \big)^2} \bigg) \bigg[\delta_{jq}+ \Big( \frac{\tau_c}{\tau_{\Omega}} \Big)^2 \omega_j\omega_q+\Big(\frac{\tau_c}{\tau_{\Omega}}\Big) \epsilon_{qjk} \omega_k \bigg] \tilde{E}_j p^q~.\label{A7}
\end{eqnarray}

 The $\delta f$ obtained in Eq.~\ref{A7} solves Eq.~\ref{R12}. Now, we will retain the label $r$ and write the current density for the hadronic species $r$ as,
\begin{equation}
 	J^{i}_{r}=-g_{r} q_{r}^2 \int \frac{d^3\Vec{p}}{(2\pi)^3}~\frac{p^{i}p^{l}}{E^2}\frac{\partial f^0_{r}}{\partial E}\Bigg(\frac{\tau_c}{1+\Big(\frac{\tau_c}{\tau_{\Omega}}\Big)^2}\Bigg)\bigg[\delta_{jl}+\Big(\frac{\tau_c}{\tau_{\Omega}}\Big)^2\omega_j\omega_l+\Big(\frac{\tau_c}{\tau_{\Omega}}\Big) \epsilon_{ljk}\omega_k\bigg]\tilde{E}_j
 	\label{A8}
\end{equation}
We can substitute the angular average, $\int d^3\Vec{p}~ p^ip^l =\int d^{3}p\left(\frac{p^2}{3}\right)\delta_{il},  (4\pi p^2dp\equiv d^{3}p)$ and the static limit ($\Vec{u}$=0) identity, $\frac{\partial f^0_{r}}{\partial E}=-\frac{f^{0}_{r}(1+af^{0}_{r})}{T}$ in Eq.~\ref{A8} to get,

\begin{equation}
 	J^{i}_{r}=\frac{g_{r} q_{r}^2}{3T} \int \frac{d^3\Vec{p}}{(2\pi)^3}~\frac{p^2}{E^2}\bigg(\frac{\tau_c}{1+\Big(\frac{\tau_c}{\tau_{\Omega}}\Big)^2}\bigg)\bigg[\delta_{ij}+\Big(\frac{\tau_c}{\tau_{\Omega}}\Big) \epsilon_{ijk}\omega_k + \Big(\frac{\tau_c}{\tau_{\Omega}}\Big)^2\omega_i\omega_j\bigg]\tilde{E}_j f^{0}_{r}(1+af^{0}_{r}).
    \label{A9}
\end{equation} 
Comparing the macroscopic expression (Ohm's law) $J^i_r=\sigma^{ij}_{r}\Tilde{E}_{j}$ with the Eq.~\ref{A9} we get, 
\begin{eqnarray}
 	&&\sigma^{ij}_{r}=\frac{g_{r} q_{r}^2}{3T} \int \frac{d^3\Vec{p}}{(2\pi)^3}\frac{p^2}{E^2}\bigg(\frac{\tau_c}{1+\Big(\frac{\tau_c}{\tau_{\Omega}}\Big)^2}\bigg)\bigg[\delta_{ij}+\Big(\frac{\tau_c}{\tau_{\Omega}}\Big) \epsilon_{ijk}\omega_k + \Big(\frac{\tau_c}{\tau_{\Omega}}\Big)^2\omega_i\omega_j\bigg]f^{0}_{r}(1+af^{0}_{r})\nn\\
 	\implies&&\sigma^{ij}_{r}=\sigma^{0}_{r} \delta_{ij} +\sigma^{1}_{r}\epsilon_{ijk}\omega_k +\sigma^{2}_{r}\omega_i\omega_j,\nn
\end{eqnarray}
where we have,
\begin{equation}
  \sigma^{n}_r = \frac{g_{r} q_{r}^2}{3T}\int \frac{d^{3}p}{(2\pi)^3}\frac{\tau_c\big(\frac{\tau_c}{\tau_{\Omega}}\big)^n} {1+\big(\frac{\tau_c}{\tau_{\Omega}}\big)^2}\times \frac{p^2}{E^{2}}f^{0}_{r}(1+af^{0}_{r}).\label{A10}  
\end{equation}
 $\sigma^{0}_r,\sigma^{1}_r,\text{ and }\sigma^{2}_r$ are scalars that make up the conductivity tensor. The total conductivity tensor is given by, $\sigma^{ij}=\sum\limits_{r}\sigma^{ij}_{r}$. The explicit form of total conductivity tensor and scalar conductivity are, 
\begin{eqnarray}
    &&\sigma^{ij}=\sum\limits_{r}\frac{g_{r} q_{r}^2}{3T} \int \frac{d^3\Vec{p}}{(2\pi)^3}\frac{p^2}{E^2}\bigg(\frac{\tau_c}{1+\Big(\frac{\tau_c}{\tau_{\Omega}}\Big)^2}\bigg)\bigg[\delta_{ij}+\Big(\frac{\tau_c}{\tau_{\Omega}}\Big) \epsilon_{ijk}\omega_k + \Big(\frac{\tau_c}{\tau_{\Omega}}\Big)^2\omega_i\omega_j\bigg]f^{0}_{r}(1+af^{0}_{r})\nn\\
    &&\sigma^{n} =\sum\limits_{r} \frac{g_{r} q_{r}^2}{3T}\int \frac{d^{3}p}{(2\pi)^3}\frac{\tau_c\big(\frac{\tau_c}{\tau_{\Omega}}\big)^n} {1+\big(\frac{\tau_c}{\tau_{\Omega}}\big)^2}\times \frac{p^2}{E^{2}}f^{0}_{r}(1+af^{0}_{r})
\label{Ape11}
\end{eqnarray}

The total current density in the rotating HRG can also be written as, 
\begin{equation}
      \Vec{J}=\sigma^{0}\Vec{\tilde{E}}+\sigma^{1}(\Vec{\tilde{E}}\times \hat{\omega})+\sigma^{2}(\hat{\omega}\cdot\Vec{\tilde{E}})\hat{\omega}
      \label{A12}
\end{equation}
%
%
\subsection{Formulas for non-relativistic HRG}\label{ape2}
Since, present work is relativistic extension of earlier Ref.~\cite{Dwibedi:2023akm}, which was done in non-relativistic framework,
so to see their difference, we have quickly gone through a HRG model in non-relativistic framework. One can rewrite Eq.~(\ref{A11}) in non-relativistic
HRG framework as
\bea
&&\sigma^n = \frac{g q^2}{T}\int \frac{d^{3}p}{(2\pi)^3}\frac{\tau_c\big(\frac{\tau_c}{\tau_{\Omega}}\big)^n} {1+\big(\frac{\tau_c}{\tau_{\Omega}}\big)^2}\times \frac{v^2}{3}f^0(1-f^0), (\vec{v}=\vec{p}/m, \text{ and } E=p^{2}/2m)\nn\\
\implies && \sigma^n = \frac{g q^2}{3T}\int \frac{d^{3}p}{(2\pi)^3}\frac{\tau_c\big(\frac{\tau_c}{\tau_{\Omega}}\big)^n} {1+\big(\frac{\tau_c}{\tau_{\Omega}}\big)^2}\times \frac{p^2}{m^{2}}f^0(1-f^0).\label{ncon}
\eea
Eq.~\ref{ncon} is useful for calculating conductivity for a single-component non-relativistic fluid. The non-relativistic HRG expression can be obtained from Eq.~\ref{ncon} as:
\bea
&&\sigma^{n} =\sum\limits_{r} \frac{g_{r} q_{r}^2}{3T}\int \frac{d^{3}p}{(2\pi)^3}\frac{\tau_c\big(\frac{\tau_c}{\tau_{\Omega}}\big)^n} {1+\big(\frac{\tau_c}{\tau_{\Omega}}\big)^2}\times\frac{p^2}{m_{r}^{2}}f^{0}_{r}(1+af^{0}_{r}).\label{nHRG}
\eea
The formulas for parallel, perpendicular, and hall conductivity for non-relativistic HRG are given by,
\be
\sigma^{||}_{\rm NHRG}\equiv \sigma_{\rm NHRG}  = \sum_{B} \frac{g_B q_{B}^2}{3T}\int \frac{d^{3}p}{(2\pi)^3}\tau_c\times \frac{p^2}{m_{B}^{2}}f^0(1-f^0)
+ \sum_{M} \frac{g_M q_{M}^2}{3T}\int \frac{d^{3}p}{(2\pi)^3}\tau_c\times \frac{p^2}{m_{M}^{2}}f^0(1+f^0)
\label{s_NHRG}
\ee
\be
\sigma^{\perp}_{NHRG} =\sum_{B} \frac{g_B q_{B}^2}{3T}\int \frac{d^{3}p}{(2\pi)^3}\frac{\tau_c}{1+\big(\frac{\tau_c}{\tau_{\Omega}}\big)^2}\times \frac{p^2}{m_{B}^{2}}f^0(1-f^0)
+\sum_{M} \frac{g_M q_{M}^2}{3T}\int \frac{d^{3}p}{(2\pi)^3}\frac{\tau_c}{1+\big(\frac{\tau_c}{\tau_{\Omega}}\big)^2}\times \frac{p^2}{m_{M}^{2}}f^0(1+f^0)
\label{sO_NHRG}
\ee
\be
\sigma^{\times}_{NHRG} =\sum_{B} \frac{g_B q_{B}^2}{3T}\int \frac{d^{3}p}{(2\pi)^3}\frac{\tau_c\big(\frac{\tau_c}{\tau_{\Omega}}\big)}{1+\big(\frac{\tau_c}{\tau_{\Omega}}\big)^2}\times \frac{p^2}{m_{B}^{2}}f^0(1-f^0)
+\sum_{M} \frac{g_M q_{M}^2}{3T}\int \frac{d^{3}p}{(2\pi)^3}\frac{\tau_c\big(\frac{\tau_c}{\tau_{\Omega}}\big)}{1+\big(\frac{\tau_c}{\tau_{\Omega}}\big)^2}\times \frac{p^2}{m_{M}^{2}}f^0(1+f^0)
\label{sOH_NHRG}
\ee
The relaxation time of any non-relativistic hadron can be written as
\be
\tau_{c}=1/(n_{NHRG} v^{NH}_{\rm av}\pi a^2)~,
\label{tc_NHRG}
\ee 
where hard sphere cross-section $\pi a^2$ is considered for hadron, having
average velocity
\be 
v^{NH}_{\rm av}=\int \frac{d^3p}{(2\pi)^3}\frac{p}{m} f_0\Big/\int \frac{d^3p}{(2\pi)^3} f_0~.\label{Nv_av}
\ee 
Each hadron will face the entire density of the system
\be
n_{NHRG}=\sum_{B}g_B\int_{0}^{\infty}\frac{d^3p}{(2\pi)^3}\frac{1}{e^{E/T}+1} +\sum_{M} g_M\int_{0}^{\infty}\frac{d^3p}{(2\pi)^3}\frac{1}{e^{E/T}-1}, (E=p^{2}/2m) 
\label{n_NHRG}
\ee
where $g_B$ and $g_M$ are Baryon and Meson spin degeneracy factors, respectively.

\bibliography{HrgEL}

\end{document}